\documentclass{llncs}
\usepackage[T1]{fontenc}
\usepackage{hyperref}
\usepackage{color}

\usepackage[utf8]{inputenc}
\DeclareUnicodeCharacter{039D}{\ensuremath{\mathrm{N}}}

\usepackage{graphicx}
\usepackage{thmtools}
\usepackage{thm-restate}
\usepackage{enumitem}
\usepackage{booktabs}
\usepackage{wrapfig}

\usepackage{graphicx}
\usepackage{epsfig}
\usepackage{xcolor}

\usepackage{subcaption}

\usepackage{cite}

\usepackage{amsmath}

\definecolor{chighlight}{named}{black}

\usepackage{xspace}

\usepackage[labelfont={bf}, font={footnotesize}]{caption}

\usepackage{enumitem}
\setlist{topsep=0pt, leftmargin=*}

\def\P#1{\mathrm{\bf P}\left\{#1\right\}}	
\def\E#1{\mathrm{\bf E}\left[#1\right]}		

\author{Ziyue Qiu \and Juncheng Yang \and Mor Harchol-Balter\thanks{Supported by NSF-CIF-2403194, NSF-III-2322973, and NSF-CMMI-2307008. } }
\institute{Carnegie Mellon University}
\begin{document}
\title{Can Increasing the Hit Ratio Hurt Cache Throughput?}

\maketitle

\begin{abstract}
Software caches are an intrinsic component of almost every computer system. 
Consequently, caching algorithms, particularly eviction policies, are the topic of many papers.  Almost all these prior papers evaluate the caching algorithm based on its {\em hit ratio}, namely the fraction of requests that are found in the cache, as opposed to 
disk.  The ``hit ratio" is viewed as a proxy for traditional performance metrics like system throughput or response time.  Intuitively it makes sense that higher hit ratio should lead to higher throughput (and lower response time), since more requests are found in the cache (low access time) as opposed to the disk (high access time). 

This paper challenges this intuition. We show that increasing the hit ratio can actually {\em hurt} the throughput (and response time) for many caching algorithms.  Our investigation follows a three-pronged approach involving {\em (i)} queueing modeling and analysis, {\em (ii)} simulation to validate the accuracy of the queueing model, and {\em (iii)} implementation and measurement. We also show that the phenomenon of decreasing throughput at higher hit ratios is likely to be more pronounced in future systems, where the trend is towards faster disks and more cores per CPU.  

\keywords{caches, hit ratio, performance evaluation, scalability, modification analysis, queueing theory, LRU, eviction policies}
\end{abstract}

\section{Introduction}
\label{s:intro}

DRAM-based software caches are widely deployed in today's infrastructure.   Examples range from simple and small page caches in laptops and mobile phones to large multi-layer distributed and heterogeneous key-value caches and object caches in the data centers, e.g., Meta Cachelib~\cite{berg_cachelib_2020}, Google CliqueMap~\cite{singhvi_cliquemap_2021}, and include many types of caches in between~\cite{borst2010distributed,willick1993disk,almeida2000hybrid}. 

The purpose of the cache is to allow fast data access.
Typically, cached items can be accessed anywhere from 100 to 10,000 times faster than those on disk~\cite{latencyNumber1}. 
The principle of caching is very simple: Store items that are likely to be accessed soon in the cache.  Store everything else on disk.

\begin{table*}[ht!]
\footnotesize
\centering
\caption{Table shows the algorithms we evaluated.  Detailed descriptions of the algorithms are in Sec.~\ref{s:models}. A classification of 12 additional algorithms is given in Table~\ref{tab:alg2}.}
\label{table:alg}
\begin{tabular}{@{}lllll@{}}
\toprule
Algorithm & Description &  \begin{tabular}[c]{@{}l@{}}  Production \\System \end{tabular} & \begin{tabular}[c]{@{}l@{}}   \underline{Our Findings}: \\ Does increasing \\ the hit ratio \\ always help? \end{tabular} & \\ \midrule \hline
LRU               & \begin{tabular}[c]{@{}l@{}} Accessed item is moved to \\front  of queue.   Evict the \\item at end of queue.  
\end{tabular} \hspace{.2in} &  \begin{tabular}[c]{@{}l@{}} Alluxio~\cite{alluxio}, \\RocksDB~\cite{rocksdb}, \\ LevelDB~\cite{leveldb} \end{tabular} &  \qquad no &  \\ \hline
FIFO              & Evicts the oldest item.               & ATS~\cite{ats}         & \qquad yes &  \\  \hline
Probabilistic LRU & \begin{tabular}[c]{@{}l@{}} Only moves accessed item \\to the head of queue with \\ some probability $1-q$. \end{tabular} & HHVM~\cite{HHVM} &     \quad depends  on $q$  & \\ \hline
\begin{tabular}[c]{@{}l@{}} FIFO-Reinsertion \\ a.k.a. CLOCK \end{tabular} &  \begin{tabular}[c]{@{}l@{}} Item at end of queue gets \\  a second chance through \\the  queue before eviction.    \end{tabular}                                                   & RocksDB~\cite{rocksdb} & \qquad yes &  \\ \hline
Segmented LRU     & \begin{tabular}[c]{@{}l@{}} Uses two LRU queues to \\differentiate items that \\ have been accessed twice. \end{tabular}                                      &  \begin{tabular}[c]{@{}l@{}} Linux~\cite{kernel} \end{tabular} & \qquad no & \\ \hline
S3-FIFO~\cite{S3FIFO}       & \begin{tabular}[c]{@{}l@{}} Uses a small FIFO queue to \\ evict most new and \\unpopular objects. \end{tabular}  & RedPanda    & \qquad yes     &  
\\ \bottomrule
\end{tabular}
\end{table*}

Utilizing a cache
improves {\em throughput}, the average number of requests served per second (RPS). This is particularly important in data processing applications where the goal is to process as many data requests as possible per unit time. 
Examples include the caches used in big-data systems (Hadoop, HDFS~\cite{hdfs}, Alluxio~\cite{alluxio}), deep learning systems (Pytorch~\cite{Pytorch}), and databases (RocksDB~\cite{dong_rocksdb_2021}).

\subsection{Cache eviction algorithms}

All caches have a common component: the cache \emph{eviction algorithm}. The eviction algorithm decides which item to evict when the cache is full. 
The \emph{most common} cache eviction algorithm is Least-Recently-Used (LRU)~\cite{oneil_lru-k_1993,eytan_its_2020,LRU-in-redis,LRU-in-cachelib}, which evicts the least recently \emph{accessed} item in the cache.
Because of its popularity, this paper will focus on LRU cache eviction.
LRU is widely used because data access patterns often show locality where recently used data have a higher chance to get reused~\cite{denning_working_1968,denning_working_1980}. 
Another common algorithm is First-In-First-Out (FIFO), which evicts the least recently \emph{inserted} item, i.e., the oldest item. 
Many more advanced eviction algorithms also exist~\cite{megiddo_arc_2003,jiang_lirs_2002,einziger_tinylfu_2017,johnson_2q_1994,song_learning_2020,beckmann_lhd_2018,vietri_driving_2018,rodriguez_learning_2021,song_halp_2023,yang_fifo_2023,einziger_lightweight_2022,
zhong_lirs2_2021,donghee_lee_lrfu_2001,kirilin_rl-cache_2019,smaragdakis_eelru_1999,ananthanarayanan_pacman_2012,berger_robinhood_2018}.

\subsection{The quest for higher hit ratio}  

The overall {\em performance goal} of a cache is to improve the request throughput and reduce the request latency.  Despite this, researchers have resorted to using the cache {\em hit ratio} (fraction of requests found in the cache) as a {\em proxy} for measuring performance~\cite{yang_large_2020,yang_segcache_2021,sabnis_tragen_2021,cheng_take_2023,mcallister_kangaroo_2021,berg_cachelib_2020,sundarrajan_footprint_2017,sabnis_jedi_2022,willick1993disk,chen1995write,zhu2012saving,hafeez2018elmem}.
Maximizing the hit ratio makes intuitive sense for improving system performance since we want to maximize the fraction of accesses that can be completed quickly from the cache and minimize the fraction of accesses that need to go to the slow disk.
\begin{quote}
{\em But what if this intuition is wrong?  What if increasing the hit ratio actually hurts performance?}
\end{quote}
This is the question investigated in this paper. The cache eviction algorithms we evaluate are summarized in Table~\ref{table:alg}.

\subsection{A 3-pronged approach to determining if higher hit ratio helps}

We take a three-pronged approach to determine if a higher hit ratio, in fact, improves throughput.

\vspace{-.1in}

\subsubsection{A. Queueing model for upper bounding throughput}
While many papers have analyzed the cache {\em hit ratio}, e.g., ~\cite{Berger15,CasaleGast21,GastVanHoudt15,CarlssonEager17,BenAmmar19,Kurose_2014,dan_towsley_1990,Rosensweig_2010,Gast_Houdt_2017,che_2002,Fricker_2012,Fofack_2014,psaras_2011_modelling,Tarnoi_2015,Garetto_2016},
the question of how the hit ratio affects throughput has been overlooked.  Perhaps this is because it seems so obvious that increasing the hit ratio can only help. 

\begin{figure}[t]
    \centering
    \includegraphics[width=0.53\linewidth]{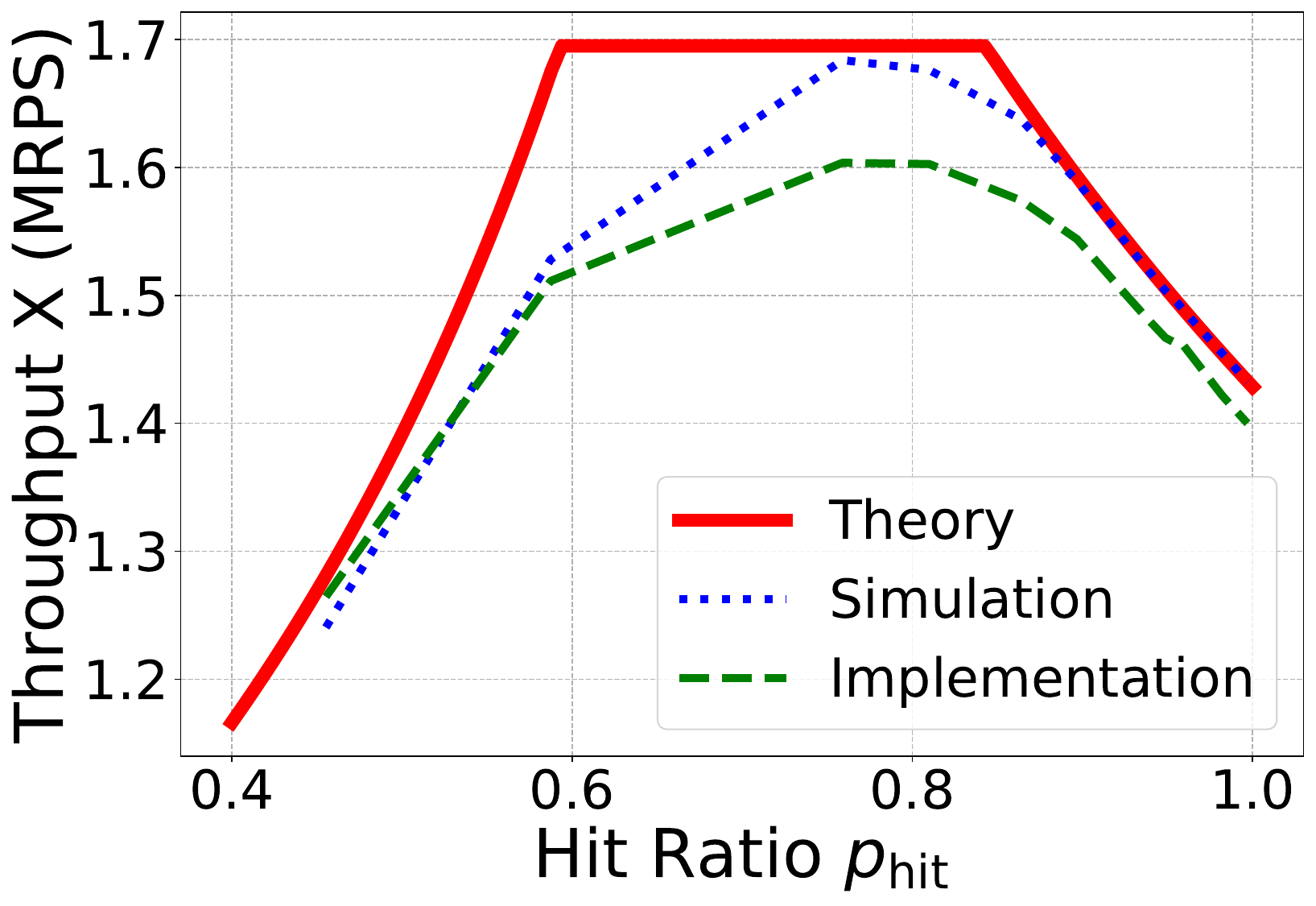}
    \caption{Throughput under LRU (measured in millions of requests per second) increases as the hit ratio increases initially but then drops when the hit ratio gets high.\vspace{-1.6em}}
    \label{fig:diagram}
\end{figure}

Instead we develop a queueing model of our cache for a range of popular eviction policies, based on measurements from our implementation. We then use queueing theory to derive an \emph{upper bound on the throughput} of this queueing model. For example, our  bound for the case of LRU is given in Figure~\ref{fig:diagram} via the red solid line. The analytic upper bound clearly shows that throughput first increases with hit ratio, then levels off, and then decreases.

\vspace{-.05in}
\subsubsection{B. Simulation of the queueing model}
Because exact analysis of the queueing model is not possible,
we next simulate the queueing model to obtain its exact throughput. For LRU, the result is shown in Figure~\ref{fig:diagram} via the blue dotted line.

\subsubsection{C. Implementation of the caching system}
Finally we implement our caching system with a range of eviction algorithms. Our prototype builds on Meta's HHVM Cache~\cite{HHVM,HHVM_Cache}, but we add a feature to emulate varying disk speeds and Multi-Programming Limits. Our results generalize to many other in-memory caches, e.g.,  CacheLib~\cite{berg_cachelib_2020}, Memcached~\cite{Memcached}, Intel OCF~\cite{Intel_OCF}, BCache~\cite{bcache}, and RocksDB LRU Cache~\cite{rocksdb}.
For LRU, the result of our implementation is shown via the green dashed line in Figure~\ref{fig:diagram}.
Importantly, the simulation result is within 5\% of the implementation result.
This tells us the performance predicted by our queueing model provides a very good estimate of system performance.

\subsection{Why increasing the hit ratio can hurt, in brief}
\label{ss:why}

The traditional intuition favoring high hit ratios goes like this:   Every request goes to the cache, but only some fraction (the misses) additionally  go to the disk.   Hence, by reducing the miss ratio, we reduce the request latency and improve the throughput and mean response time.

Queueing theory provides a careful bottleneck analysis, explaining why the above intuition can be wrong. As disks have become faster and the number of CPU cores has increased, the disk has become able to support more concurrent requests more quickly.  This has caused the bottleneck to shift from the disk access to various cache operations. While there are cache operations on both the ``hit path" and the ``miss path"  (see Figures~\ref{fig:LRUcache}, ~\ref{fig:ProbLRUcache}, and ~\ref{fig:SLRUcache}), the ``demand" created by a cache operation depends on the product of its service time and the probability that we follow that path.  Consequently, as $p_{hit}$ increases, the dominant bottleneck can shift from a cache operation on the miss path to one on the hit path.  Beyond this critical switching point, increasing $p_{hit}$ only hurts performance (throughput decreases and response time increases). As we'll see, queueing theory produces an excellent prediction of the critical $p_{hit}$ point.

\subsection{Contributions}

The contributions of this paper are summarized below:
\begin{itemize}
\item This paper shows that increasing cache hit ratio can hurt throughput for many LRU-like cache eviction algorithms. We show this via a three-pronged approach, involving queueing theory, simulation, and implementation. 
\item We develop {\em queueing models} that allow us to understand the effect of the cache hit ratio on system throughput and mean response time. This is done for six different caching policies.
The modeling is non-trivial and, to the best of our knowledge, such models do not exist in prior work. 

\item While our queueing analysis only provides upper bounds, the analysis clearly indicates that throughput initially rises with hit ratio and then drops with hit ratio. The analysis also provides the critical switching point, $p^*_{hit}$.
\item We also {\em implement} many caching policies, including LRU, FIFO, Probabilistic LRU, and CLOCK. We validate the correctness of our queueing models, by {\em simulating} the queueing models and showing that the simulation results match the implementation results within 5\% for all caching policies studied.
\item We evaluate the effect of changing disk latency as we move from older disk speeds (500 $\mu s$) to current disk speeds (100 $\mu s$) to future disk speeds (5 $\mu s$).   
As we move to future disk speeds, the critical hit ratio,  $p^*_{hit}$, after which throughput starts to deteriorate moves earlier and earlier.  
\item We also evaluate the effect of another trend, increasing the number of CPU cores (concurrency).  We find that $p^*_{hit}$ moves earlier for higher concurrency, i.e., higher Multi-Programming Level (MPL). See Figure~\ref{fig:intro:slru-simu}.
\end{itemize}

\section{Background}

{\bf DRAM-based software caches:} As explained in Section~\ref{s:intro}, DRAM-based software caches are extremely common today.
In a DRAM-based software cache, DRAM is used to cache the accessed data, making cache access time quick. This is in contrast to SSD-based software caches where the cache access is 100 times slower and far less concurrent than in the DRAM system. This paper focuses on DRAM-based software caches.  
Hardware caches are also outside the scope of this paper as are network-connected caches. 


\noindent{\bf Hardware trends:}  The performance of DRAM-based software caches is highly influenced by the number of cores on which the software runs and the backend disks.  
Before 2000, CPUs had only one core~\cite{Multi-core_Processor}, but a modern CPU has 32-192 cores~\cite{amdcpus}. The increase in CPU cores enables better performance, allowing more concurrent requests; however, it also presents challenges because the cores need to coordinate with each other.  
Today's backend disks are implemented on SSDs, where high-end SSDs have latencies around 5 $\mu s$~\cite{ssd-fast,ssd-fast1}, low-end SSDs show latencies of a few hundred $\mu s$~\cite{ssd-low1,ssd-low2}, and most commercial SSDs have latency in between~\cite{ssd-normal,ssd-normal2,ssd-normal3}. 
These backend disks support massive concurrency~\cite{ssd-parallel1,ssd-parallel2}, allowing requests at the disk to all be served in parallel without queueing.  










\section{A three-pronged approach to LRU}
\label{s:lru}
\vspace{-.05in}
Because LRU is the most common caching policy, we devote this section to LRU's performance. Our purpose is to explain why increasing the hit ratio can lead to decreased throughput. 
The numbers used in the queueing model are measured from our prototype built on Meta's DRAM-based software cache, with hardware to be described in Section~\ref{ss:implementation}. Takeaways are summarized in Section~\ref{ss:takeaway}.

\vspace{-.07in}
\subsection{A closed-loop queueing model of LRU}
\label{ss:closed-loop-lru}
\vspace{-.1in}
As with many caching systems evaluations and benchmarks, e.g., Cachelib~\cite{berg_cachelib_2020}, YCSB benchmark~\cite{cooper_benchmarking_2010}, S3-FIFO~\cite{S3FIFO}, and FrozenHot~\cite{qiu_frozenhot_2023}, our system is best modeled by a   {\em closed-loop queueing model},
where new requests are triggered by the completion of previous requests.  There is a fixed Multi-Programming Limit (MPL), $N$, denoting the number of requests that can be in the system at a time (this is dictated by the number of cores -- in our case 72). 

\vspace{-1em}
\subsubsection{Modeling concurrency in a closed-loop model}

Each request is handled by a single core. The total number of requests in the system is thus limited by the total number of cores. Throughout, we assume that there is one CPU with 72 cores, and thus we can process 72 requests concurrently.

We thus model our caching system via a closed-loop queueing model, where a new request is allowed to enter only when some other request completes. The {\em multi-programming limit (MPL)} for the system is $N = 72$. See \cite[Chapters 2,6,7]{PerformanceModeling13} for background on modeling closed-loop queueing models.

\vspace{-.1in}
\subsubsection{Modeling disk access and cache access}
\label{sec:modelingdiskcache}

Our disk has enough concurrency that it can be accessed simultaneously by all 72 requests. Thus, in queueing speak, we can model the disk as a {\em think station} (infinite number of simultaneous service stations) with mean think time $\E{Z_{disk}}= 100 \mu s$. 
Likewise, the cache lookup can also be executed concurrently. Thus, the cache lookup can also be modeled as a think station but with a much faster mean think time of $\E{Z_{cache}} = 0.51 \mu s$. Note that a think station is different from a queue station in that there is no queueing at a think station -- every request starts running immediately, and requests are served concurrently.

\vspace{-.1in}
\subsubsection{Modeling software global list operations}
\label{sss:model-ES}
In an {\em LRU} cache, all cached items are stored in a single {\em global linked list}, 
where the least-recently-used item is the ``tail'' item and is the one to evict, while the most-recently-used item is at the ``head'' of the list. 
A request for some item $d$ first looks for $d$ in the cache. Either  $d$ is in the cache (called a ``hit”), or it is not (a ``miss”). The probability of a hit is denoted by $p_{hit}$ and the probability of a miss is $p_{miss}$, where $p_{hit} + p_{miss} = 1. $

If the request is a hit, then two things need to happen:
\begin{enumerate}
\item The item $d$ must be delinked from its position in the global linked list. This is the {\em delink operation}. The delink time is denoted by the service time random variable $S_{delink}$. 
\item The item $d$ needs to be attached to the head of the global linked list. This is called the {\em cache head update}. The head update time is denoted by the random variable $S_{head}$. 
\end{enumerate}

The actual ``reading time'' (the time to read a 4 KB block from DRAM)  is not included in our model. The reason is that this is very small compared with cache lookup, is handled concurrently,  and is the same across all algorithms.

If the request is a miss, then three things need to happen:
\begin{enumerate}
\item The item $d$ needs to be found on disk. 
\item The least-recently-used item, at the tail of the global list, needs to be removed. We call this the {\em cache tail update} and denote it by the random variable $S_{tail}$. 
\item Item $d$ needs to be attached to the head of the global linked list. This is the {\em cache head update} mentioned earlier, denoted by random variable $S_{head}$.
\end{enumerate}

\begin{figure}[t]
\centering
\includegraphics[width=0.64\linewidth]{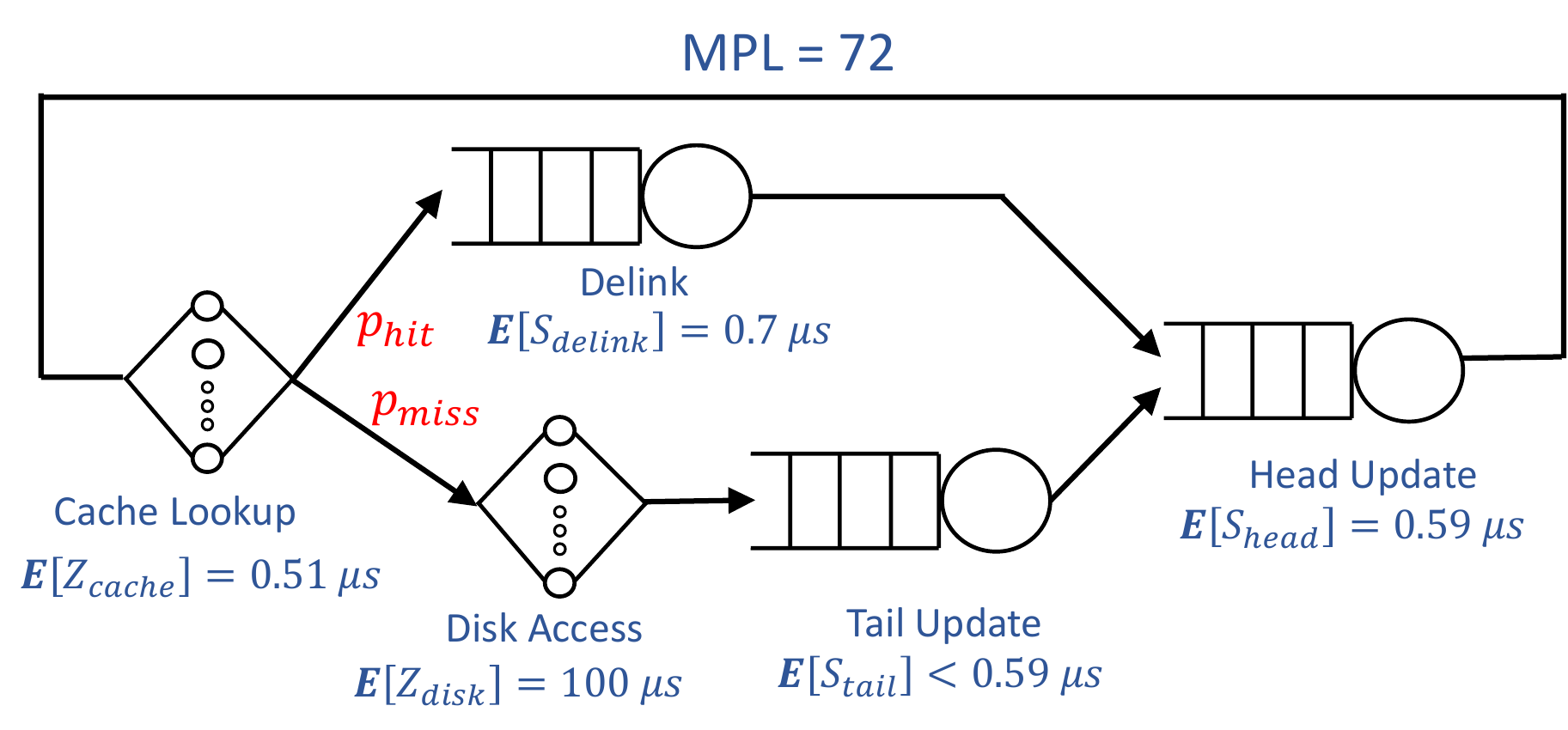}
\vspace{-1.2em}
\caption{Queueing model of LRU cache. \vspace{-1.6em}}
\label{fig:LRUcache}
\end{figure}

We can thus model the LRU caching system via the queueing model shown in Figure~\ref{fig:LRUcache}.

{\color{chighlight} Understanding how to properly model $S_{head}$, $S_{tail}$, and $S_{delink}$ is important and non-trivial and will come up again when discussing different eviction policies. We will explain these here. 
We start with $S_{head}$, which denotes the time needed for a head update on the global linked list. This has {\em two components}. The first component is obvious: just the update to the global list. The second component is less obvious: communicating this update to all the other requests in the queue. Specifically, the core performing the head update needs to communicate with all the other cores in the head update queue to alert them that this head update is happening. Obviously, if there are many requests in the head update queue, then this communication will take longer.

In summary, $S_{head}$ consists of a {\em constant} head update time and a communication time dependent on {\em queue length}.
When the head update is the bottleneck operation, our measurements show that these two components add up to $\E{S_{head}} = 0.59 \mu s$. When the head update is not the bottleneck operation, the queue length at the head update queue drops, thus $\E{S_{head}}$ decreases. However, as we'll see in the analysis of LRU (Section~3.2), in the case where an operation is not a bottleneck, its service time does not impact overall throughput much. Specifically, any value in $0 < \E{S_{head}} < 0.59 \mu s$ will produce the same throughput if a head update is not the bottleneck operation.  We find that $S_{head}$ follows approximately a Bounded Pareto distribution (with $\alpha = 0.45$) ranging from $0.1 \mu s $ to $1.2 \mu s$. However, as we'll see in Section~3.2, the throughput analysis of the queueing model is only influenced by the mean, $\E{S_{head}}$. Similarly, when the delink queue is the bottleneck, $\E{S_{delink}} = 0.7 \mu s$. When the delink queue is not the bottleneck, its value has little impact on analysis.

{\em Importantly}, when we look at other eviction algorithms, the number for $\E{S_{head}}$ can change because the queue length at the head update queue will change.
To measure $\E{S_{head}}$, we create a setting where the head update is the bottleneck operation. This means that the head update queue will be flooded.  Consequently the service time, $S_{head}$, is simply the inter-departure time from this flooded queue (namely the time between consecutive departures from the queue), which is easy to measure.  Likewise, it is easy to measure $\E{S_{delink}}$, because that device can also be made into the bottleneck (by setting $p_{hit}$ appropriately).  

The case of $S_{tail}$ is slightly different because the tail update is {\em never} the bottleneck operation. This makes it difficult to precisely measure $\E{S_{tail}}$ because we can't keep the tail update server fully utilized as we were able to do for the other servers. Fortunately, again, as shown in Section~3.2, the precise value will not matter.
}


\subsection{Analysis of LRU queueing model}
\label{ss:analysis}

The goal of this section is to determine analytically how the system throughput is affected by hit ratio, $p_{hit}$. We first illustrate our analysis assuming that mean disk latency is $\E{Z_{disk}} = 100 \mu s$ and then generalize to other disk latencies. Our analysis of closed systems is based on \cite[Chapters 6,7]{PerformanceModeling13}) and produces an upper bound on throughput for the queueing network in Figure~\ref{fig:LRUcache}.

The first step is to determine the {\em mean think time} of the system, $\E{Z}$, where $\E{Z}$ is the mean time spent on accesses that can be executed concurrently by all cores.  This includes cache lookup and disk access:
\begin{eqnarray*}
\E{Z} & = & \E{Z_{cache}} + (1 - p_{hit}) \cdot \E{Z_{disk}} \\
&= & 0.51 + (1 - p_{hit}) \cdot 100 = 100.51 - 100 p_{hit}
\end{eqnarray*}

For each queue, we now compute the {\em device demand}, which is the expected service demand on the corresponding device, per request into the system.  The device demands are:
  \begin{eqnarray*}
D_{delink} & = & p_{hit} \cdot 0.7 \\
D_{tail} & < & (1 - p_{hit})  \cdot 0.59 \\
D_{head} & = & 0.59
    \end{eqnarray*}


The {\em total demand}, $D$, is the sum of the device demands: 
\begin{eqnarray*}
  D & = &  D_{delink} + D_{tail} + D_{head}  
  \end{eqnarray*}
Because $0 < D_{tail} < 0.59$, we have upper and lower bounds on $D$ as follows:
\begin{eqnarray*}
  0.7 p_{hit} + 0.59 < & D & < 0.7 p_{hit} + 0.59 (1 - p_{hit}) + 0.59 \\
0.7 p_{hit} + 0.59 <  &D & < 0.11 p_{hit} + 1.18
\end{eqnarray*}

The next step is to determine the {\em bottleneck device}, which is the device with the highest demand.  We can see that the bottleneck device is the delink device if $p_{hit}$ is sufficiently high, specifically $p_{hit} > 0.84$.  Otherwise, the bottleneck device is the head update device.   We write this as:  
  $$D_{max} = \max\left(0.59, 0.7 p_{hit}\right) = \left\{ \begin{array}{cc} 0.59 & \ \mbox{ if } p_{hit} < 0.84 \\ 0.7 p_{hit} & \ \mbox{ if } p_{hit} > 0.84 \end{array} \right. $$
  
We use $X$ to denote system throughput.  From \cite[Theorem 7.1]{PerformanceModeling13}, we know that  $X$ is upper-bounded by two terms, as follows:    
$$X \leq \min\left( \frac{N}{D + \E{Z}}  \ , \  \frac{1}{D_{max}} \right). $$
Substituting in the expressions for $\E{Z}$, and $D_{max}$ that we have already derived, as well as the lower bound on $D$ and the fact that $N=MPL = 72$, we have that, for the case of $\E{Z_{disk}}= 100 \mu s$:

\begin{eqnarray}
X_{\mbox{\tiny LRU}} \leq \min\left( \frac{72}{101.1 - 99.3 p_{hit}}\ , \ \frac{1}{\max(0.59, 0.7p_{hit})} \right)
\label{eqn:XLRU100}
\end{eqnarray}

\begin{figure*}[!ht]
\centering
\begin{subfigure}{.33\textwidth}
  \centering
  \includegraphics[height=1.in]{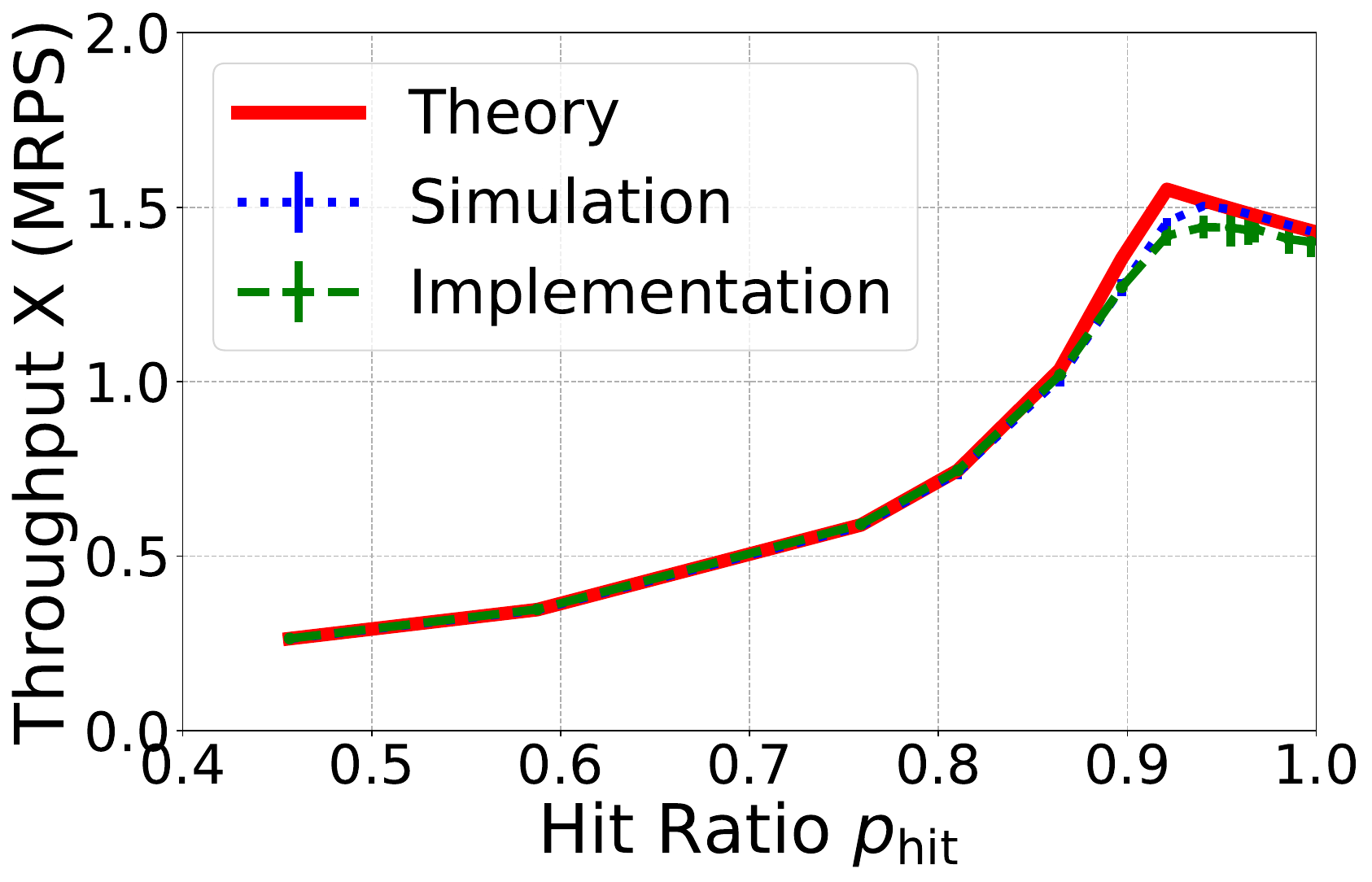}
  \caption{Disk latency 500 $\mu$s}
  \label{fig:lru:sub1}
\end{subfigure}%
\begin{subfigure}{.33\textwidth}
  \centering
  \includegraphics[height=1.in]{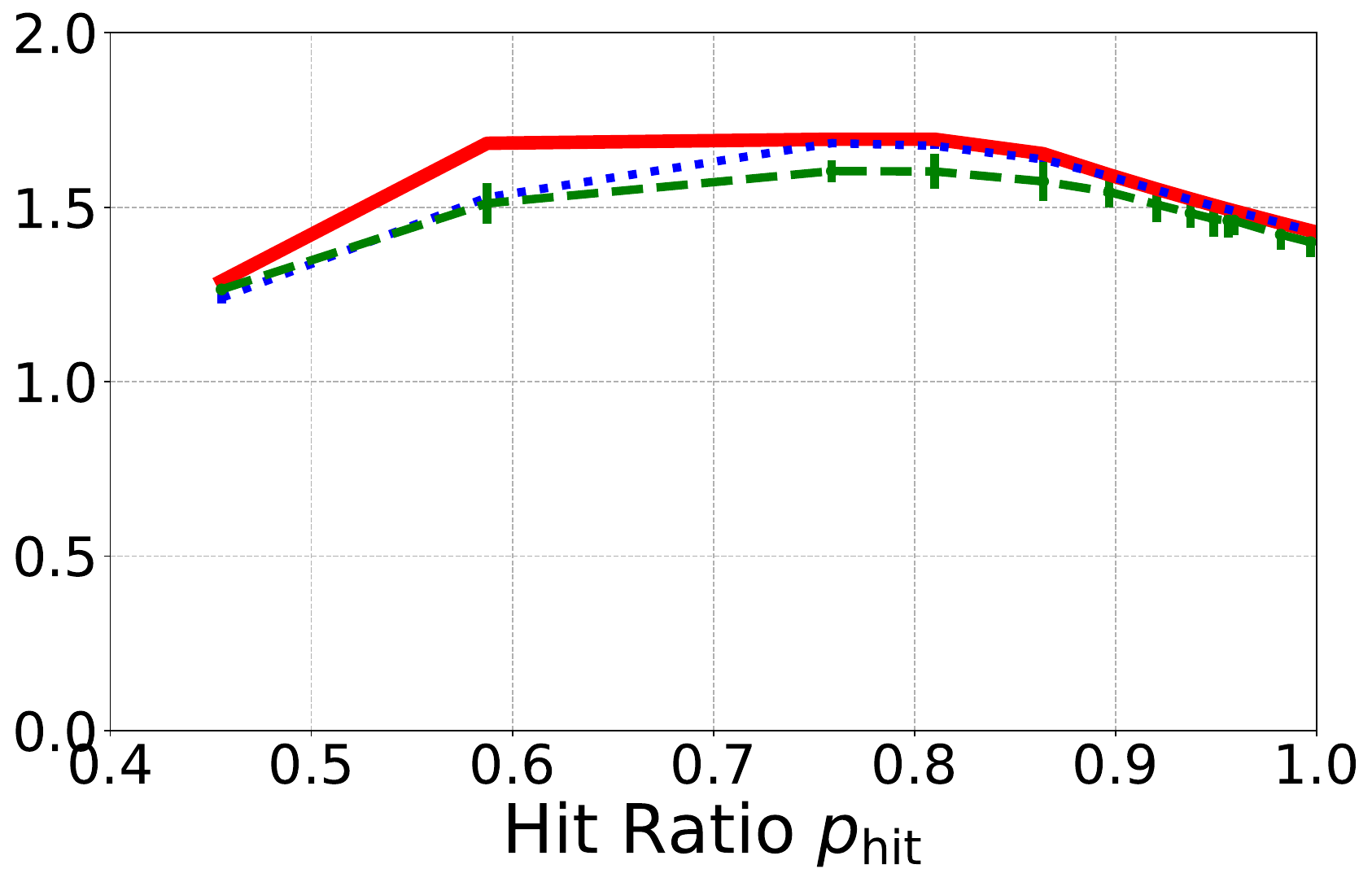}
  \caption{Disk latency 100 $\mu$s}
  \label{fig:lru:sub2}
\end{subfigure}
\begin{subfigure}{.33\textwidth}
  \centering
  \includegraphics[height=1.in]{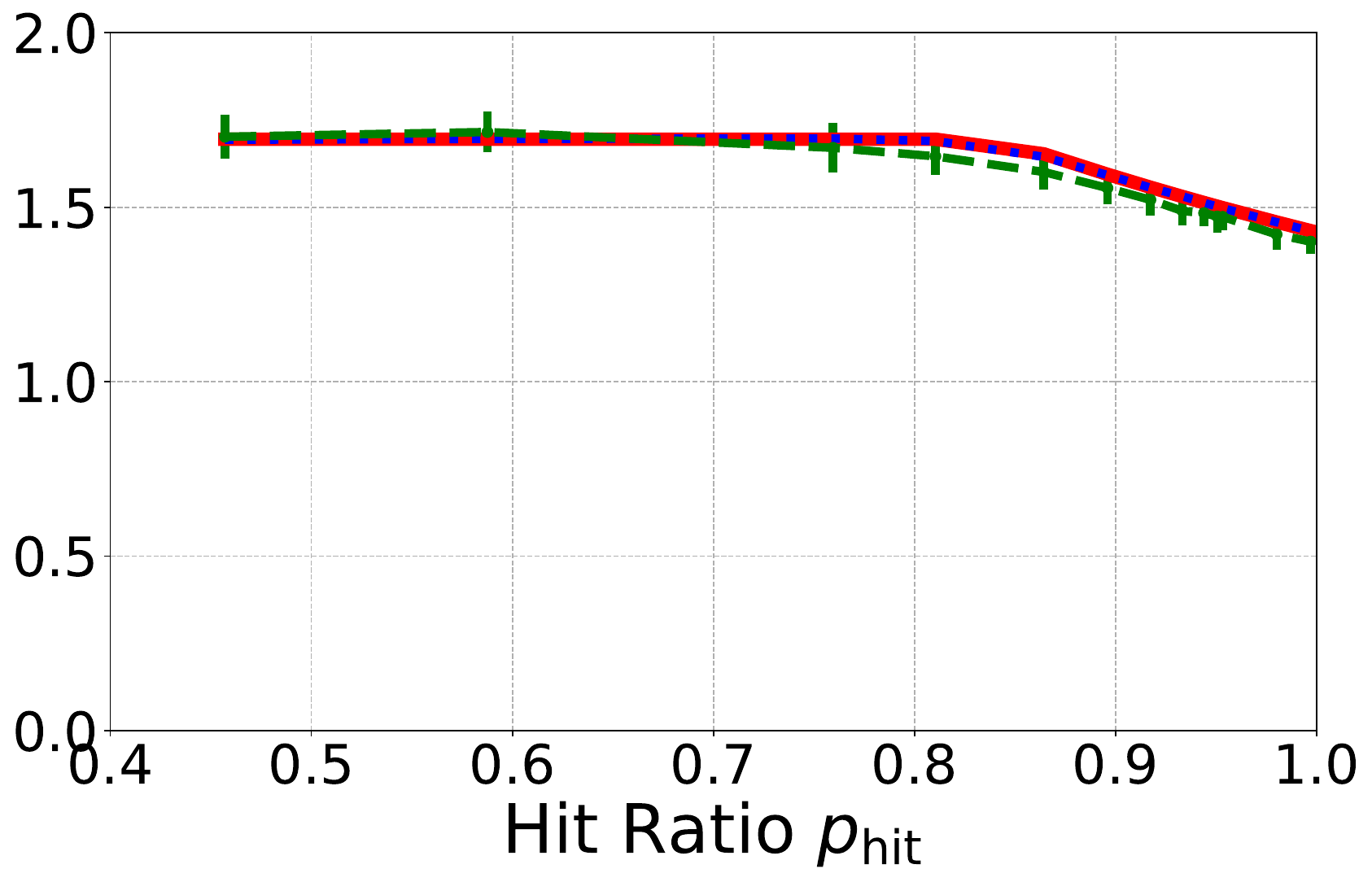}
  \caption{Disk latency 5 $\mu$s}
  \label{fig:lru:sub3}
\end{subfigure}
\vspace{-1.6em}
\caption{Results for theory, implementation, and simulation under an LRU cache.   The throughput of the LRU cache decreases at higher hit ratios.  This trend becomes more pronounced as we move towards lower disk latencies -- from (a) to (b) to (c). \vspace{-1.2em}}
\label{fig:lru:LRU-simu}
\end{figure*}

Equation (\ref{eqn:XLRU100}) represents an {\em upper bound on throughput}, shown in red in Figure~\ref{fig:lru:LRU-simu}(b). This turns out to be a very good bound on the measured throughput from our implementation. 
When $p_{hit} < 0.59$, the first term in (\ref{eqn:XLRU100}) is the relevant bound (minimum term).   When $ 0.59 < p_{hit} < 0.84$, the second term in (\ref{eqn:XLRU100}) is the relevant bound, where the max term in the denominator is $0.59$.  When $p_{hit} > 0.84$, the second term in (\ref{eqn:XLRU100}) is again the relevant bound, but the max term in the denominator is $0.7p_{hit}$. 

Recall that $0 < \E{S_{tail}} < 0.59$.  In the above analysis, we assumed that $\E{S_{tail}} = 0$ because we wanted an upper bound on $X$.  If instead, we had used any value of $\E{S_{tail}}$ in the range between $0$ and $0.59$, the impact on our result for $X$ would be very small ($< 0.5\%$).  The reason is that changing $\E{S_{tail}}$ would only change $D$, not $D_{max}$. Hence only the first term in (\ref{eqn:XLRU100}) would change, and we only care about this first term when $p_{hit}$ is low, specifically $p_{hit} < 0.59$.   Within this first term, $D + \E{Z}$ would change from $101.1  - 99.3 p_{hit}$ to $101.69  - 99.89 p_{hit}$. 

The above analysis assumed that $\E{Z_{disk}}= 100 \mu s$.  If we redo the analysis for the case where $\E{Z_{disk}} = 5 \mu s$, the throughput is (\ref{eqn:XLRU5}), as shown in red in Figure~\ref{fig:lru:LRU-simu}(c).

\begin{eqnarray}
X_{\mbox{\tiny LRU}} \leq  \min\left( \frac{72}{6.1  - 4.3 p_{hit} } \ , \ \frac{1}{\max(0.59, 0.7p_{hit})} \right)
\label{eqn:XLRU5}
\end{eqnarray}

If we repeat the analysis for the case where $\E{Z_{disk}} = 500 \mu s$, we get (\ref{eqn:XLRU500}) shown in 
red in Figure~\ref{fig:lru:LRU-simu}(a).

\begin{eqnarray}
X_{\mbox{\tiny LRU}} \leq  \min\left( \frac{72}{501.1  - 499.3 p_{hit} } \ , \ \frac{1}{\max(0.59, 0.7p_{hit})} \right)
\label{eqn:XLRU500}
\end{eqnarray}

\subsubsection{Discussion of the analytic results.}
\label{sss:intuition}

We've seen that increasing the hit ratio leads to lower throughput when the hit ratio is high.  We have shown via a queueing analysis why this happens. From a more intuitive perspective, when the hit ratio is high, we see that the {\em delink} operation becomes the bottleneck. Hence almost all requests are queued behind the delink server in Figure~\ref{fig:LRUcache}.  Thus,  while it seems that we are saving time by not going to disk, we instead are wasting time by having to queue up at the delink device.  Thus requests can actually take {\em longer}. This longer response time translates to a drop in throughput.

In all the curves of Figure~\ref{fig:lru:LRU-simu}, we find that there is some point,  $p^*_{hit}$, after which increasing the hit ratio only hurts.  This point $p^*_{hit}$ {\em decreases} as the mean disk latency decreases.  
Thus our message about not blindly increasing the hit ratio will become more and more valid as we move to faster disks.

Throughout, we have looked at throughput, but we could instead have looked at {\em mean latency}, namely the time from when a request is submitted until it completes.  Given that we have a closed-loop setting, mean response time and throughput are inversely related (see \cite[Chapters 6,7]{PerformanceModeling13}). Hence mean response time {\em increases} for higher hit ratios.     



\subsection{Simulation evaluation}
\label{ss:simulation}

Recall that our queueing analysis from Section~\ref{ss:analysis} only provides upper bounds on throughput.
To get the exact throughput of the queueing network, we turn to simulation.
We use an event-driven simulation\footnote{Open-sourced: \url{https://github.com/ziyueqiu/CacheThputSim.git}} based on our measurements of the device latencies (service times).  We note that our results appear to be  insensitive to the particular service time distributions used.  This is consistent with the findings in \cite{schroeder_open_2006} for closed-loop models.

The results of our simulation are shown in Figure~\ref{fig:lru:LRU-simu} via dotted blue lines. We also show 95\% confidence intervals but they are typically too small to be visible. As expected, the results of simulation lie below the red theory lines, which represent the theoretical upper bound.


\vspace{-1em}
\subsection{Implementation Setup and Results}
\label{ss:implementation}

In both the analysis (Section~\ref{ss:analysis}) and the simulation (Section~\ref{ss:simulation}), we were evaluating the queueing network in Figure~\ref{fig:LRUcache}. We now study LRU via our implementation that is independent of any queueing network. 
Our implementation is a prototype based on Meta's HHVM Cache.  Figure~\ref{fig:lru:LRU-simu} shows the results of our LRU implementation via green dashed lines  with 95\% confidence intervals.

{\bf Experimental setup:} Our experiments use dual-socket servers with Intel Xeon Platinum 8360Y 36-core processors (Ubuntu 20) at 2.4GHz, running on CloudLab platform~\cite{Cloudlab}. To avoid 
NUMA impacts, our evaluations only use a single socket with hyperthreading enabled. To provide consistent results, we disable turbo-boosting and fix the per-core frequency at 3.1 GHz.

Our Intel Xeon Platinum CPU allows for 72 cores to run concurrently.  This limits the number of requests that can be in the system at once to 72. Each request accesses a 4KB block of data,  the common size in database block caches. 
We emulate three different disk speeds:
$500 \mu s$, $100 \mu s$, and $5 \mu s$.  
We consider $p_{hit}$ in the range of $[0.4, 1]$, with a step size of $0.05$ in most cases, but a step size of $0.02$ for higher $p_{hit}$ values. 
Each experiment is run 20 times. 

{\bf Workload creation:}
For request generation, we employ 72 client threads, where each thread is assigned to a single CPU core. We use a synthetic popularity distribution following the Zipfian parameter $\theta$ = 0.99, representative of cache accesses from e-Commerce websites~\cite{chen_hotring_2020} and social networks~\cite{yang_large_2020}. Recognizing that our goal is to assess the impact of hit ratio on throughput, and that the popularity distribution only affects the hit ratio, we determine that it is sufficient to test with this well-established Zipfian distribution without the need for a broader range of models or real-world access traces. Throughput measurements are conducted after some warmup period, when the cache is full.

{\bf Results of implementation:}
Our implementation results and simulation results are always within $5\%$. 

\vspace{-.1in}
\subsection{Summary and takeaways}
\label{ss:takeaway}

We started the section by presenting a queueing model of our LRU caching system (Section~\ref{ss:closed-loop-lru}). We were able to derive an {\em upper bound} on throughput in our model as a function of the hit ratio (Section~\ref{ss:analysis}). Our analysis elucidated that when the hit ratio gets high, the queueing bottleneck shifts from the head update operation to the delink operation.  Increasing the hit ratio beyond $p^*_{hit}$ puts extra demand on the delink operation, resulting in longer delays and lower throughout. We next simulated the queueing network (Section~\ref{ss:simulation}), determining the exact throughput as a function of hit ratio. Finally, in Section~\ref{ss:implementation}, we implemented our LRU caching system, yielding results  within 5\% of the simulation.

There are two takeaways. First, because the implementation matches the simulation, we conclude that our queueing model is an excellent representation of the real system, at least with respect to understanding system throughput as a function of hit ratio. Second, we see that a simple queueing analysis enables us to easily predict  $p^*_{hit}$.  This foreshadows a {\em theme of this paper} -- queueing analysis alone suffices to predict the effect of hit ratio on throughput for many  policies.




\section{Evaluation of other policies}
\label{s:models}

In this section, we will apply our three-pronged approach to the remaining algorithms in Table~\ref{table:alg}. 

\subsection{FIFO}


In the {\em FIFO policy} we  store all cached objects into a single global linked list. 
The list maintains the same ordering the entire time, with new items being added to the list head and the oldest item dropping off the tail.  

If a request for item $d$ is a cache hit, {\em nothing} happens to the global linked list.  
When the request is a miss, the list must be updated.   Specifically: 
\begin{enumerate}
\item The item, $d$, needs to be read from disk.  
\item The oldest item needs to be removed from the global list, requiring a cache tail update. 
\item Item $d$ must be attached to the list head, requiring a cache head update.  
\end{enumerate}

\begin{figure}[t]
\centerline{\psfig{file=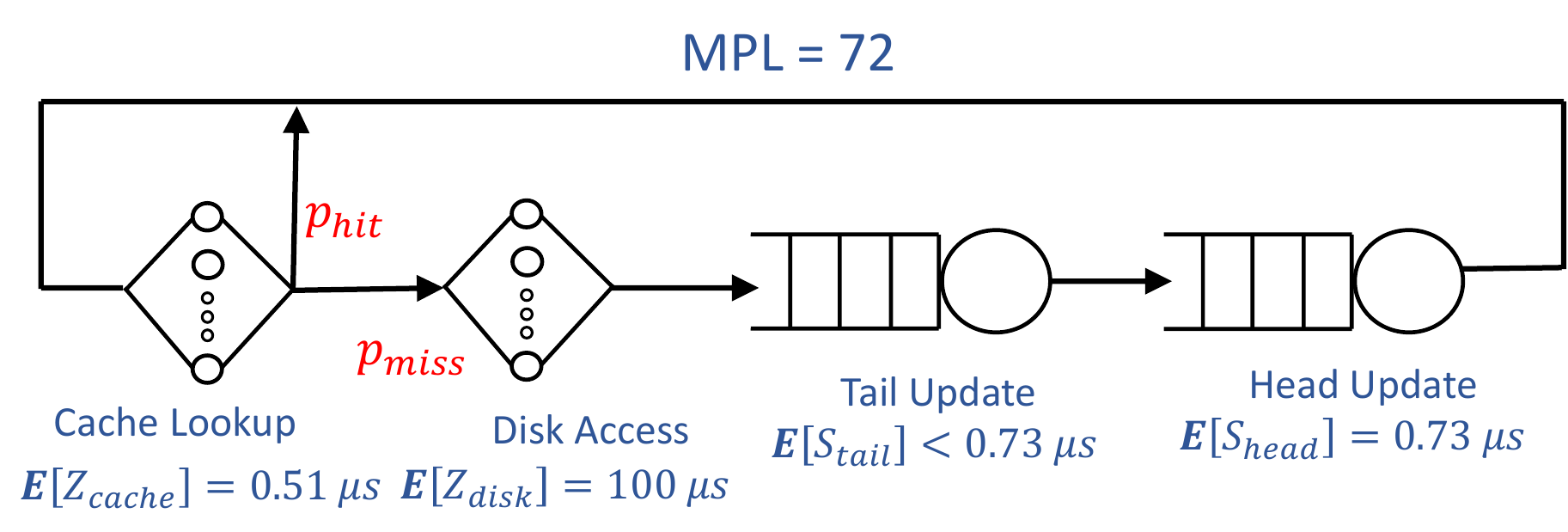,height=1.28in}}
\vspace{-1.6em}
\caption{Queueing model of FIFO cache. \vspace{-1.6em}}
\label{fig:FIFOcache}
\end{figure}

We can thus model the FIFO caching system via the closed-loop queueing model in Figure~\ref{fig:FIFOcache}.  
It may seem strange that  $\E{S_{head}}$ is higher in the FIFO system (Figure~\ref{fig:FIFOcache}) than in the LRU system (Figure~\ref{fig:LRUcache}).
To understand why this happens, recall 
that there are two components to $\E{S_{head}}$: {\em (i)} a constant time needed for a head update to the global linked list, and {\em (ii)} a communication time proportional to the queue length at the head update queue.  Now, in LRU there are 72 jobs in the system, split among {\em three} queues, but in FIFO, the 72 jobs are split among only {\em two} queues.  Consequently the expected queue length of each of FIFO's two queues is larger than each of LRU's three queues. This in turn means that component (ii) is higher under FIFO, explaining why $\E{S_{head}}$ is larger under FIFO.


We now follow the same approach that we used for LRU to analyze our closed queueing network.  

Again the {\em mean think time} of the system is: 
\begin{eqnarray*}
    \E{Z} & = & \E{Z_{cache}} + p_{miss} \cdot \E{Z_{disk}}  =  100.51 - 100 p_{hit}
\end{eqnarray*}

For each queue, we now compute the {\em device demand}: 
  \begin{eqnarray*}
D_{tail} & < & (1 - p_{hit})  \cdot 0.73  \hspace{0.3in}
D_{head}  =  (1 - p_{hit}) \cdot 0.73
    \end{eqnarray*}

The {\em total demand}, $D$, is the sum of the device demands: $$D  =   D_{tail} + D_{head} $$

Because $0 < D_{tail} < (1 - p_{hit})  \cdot 0.73$, we have upper and lower bounds on $D$ as follows:
\begin{eqnarray*}
  (1 - p_{hit}) \cdot 0.73 < & D & < (1 - p_{hit}) \cdot 0.73 + (1 - p_{hit}) \cdot 0.73 \\
0.73 - 0.73 p_{hit} <  &D & < 1.46 - 1.46 p_{hit}  
\end{eqnarray*}

The {\em bottleneck device} is always the head update, so:
  $$D_{max} = 0.73 - 0.73p_{hit}. $$
  
   Given all the above terms, where $\E{Z_{disk}}= 100 \mu s$, from \cite[Theorem 7.1]{PerformanceModeling13} our upper bound on throughput $X$ is:
\begin{eqnarray}
X_{\mbox{\tiny{FIFO}}} \leq  \min\left( \frac{72}{101.24  - 100.73 p_{hit} } \ , \ \frac{1}{0.73 - 0.73p_{hit}} \right) 
\label{eqn:XFIFO100}
\end{eqnarray}

\begin{figure*}[t]
\centering
\begin{subfigure}{.33\textwidth}
  \centering
  \includegraphics[height=1.in]{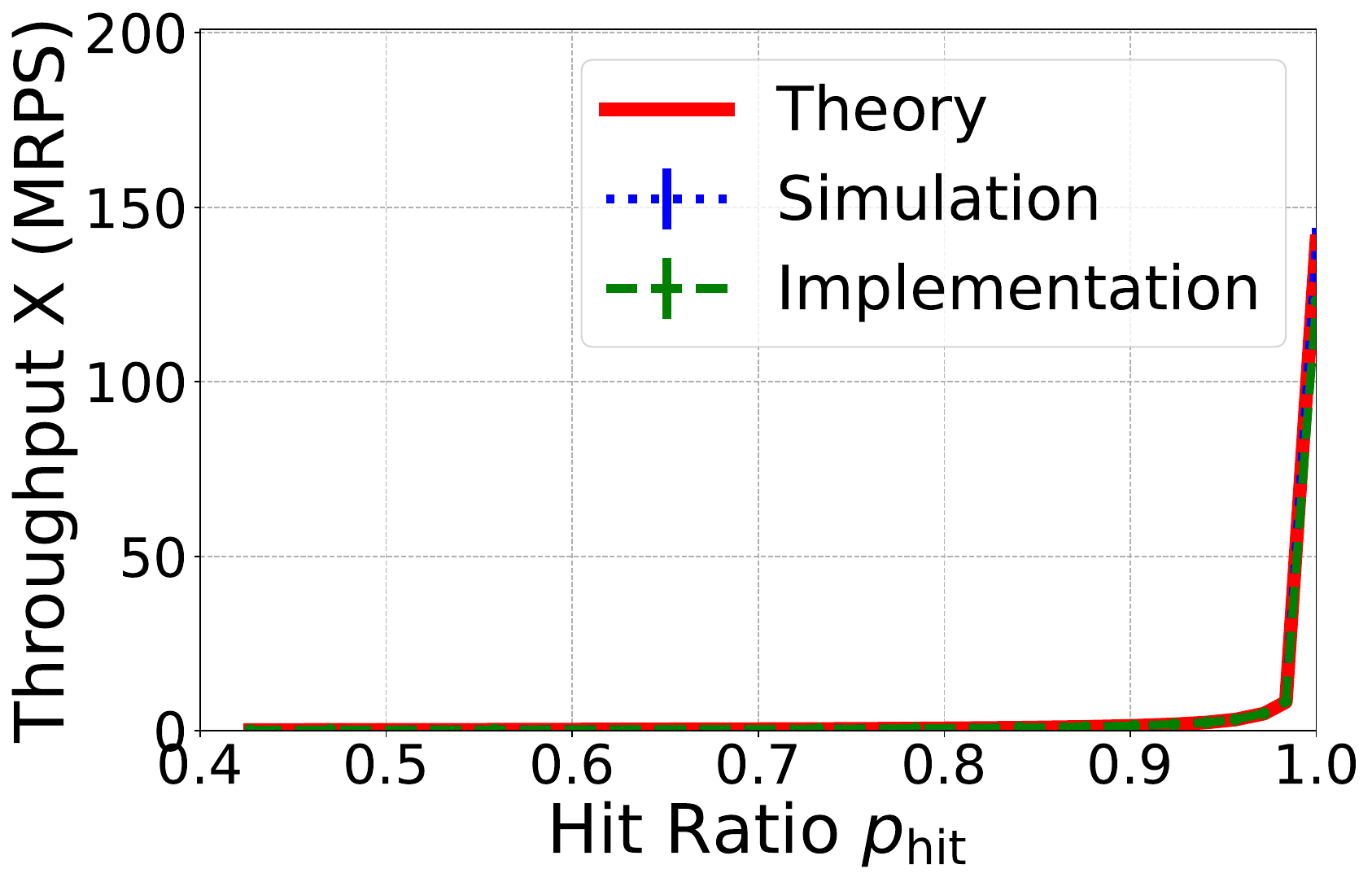}
  \caption{Disk latency 500 $\mu$s}
\end{subfigure}%
\begin{subfigure}{.33\textwidth}
  \centering
  \includegraphics[height=1.in]{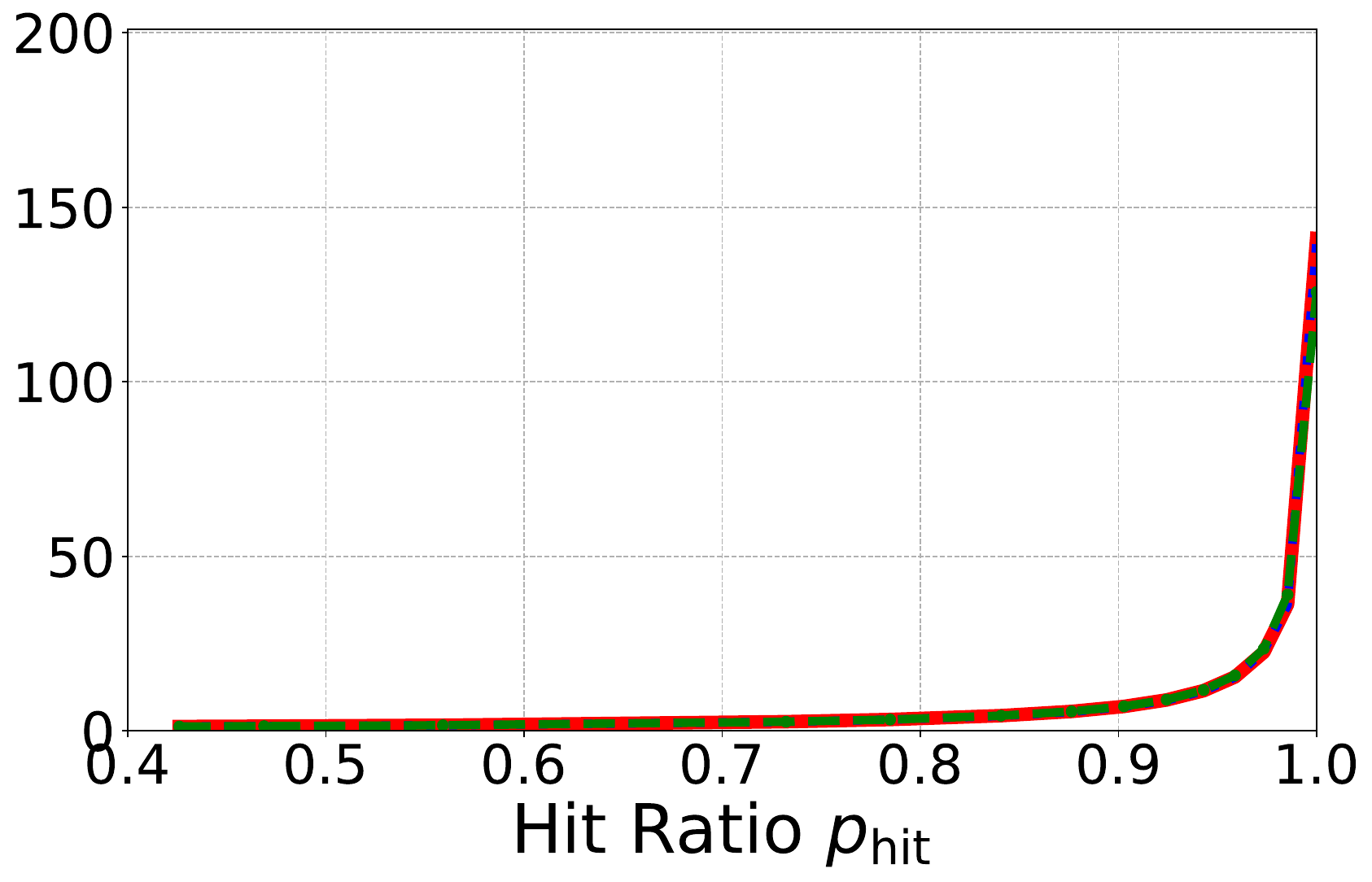}
  \caption{Disk latency 100 $\mu$s}
\end{subfigure}
\begin{subfigure}{.33\textwidth}
  \centering
  \includegraphics[height=1.in]{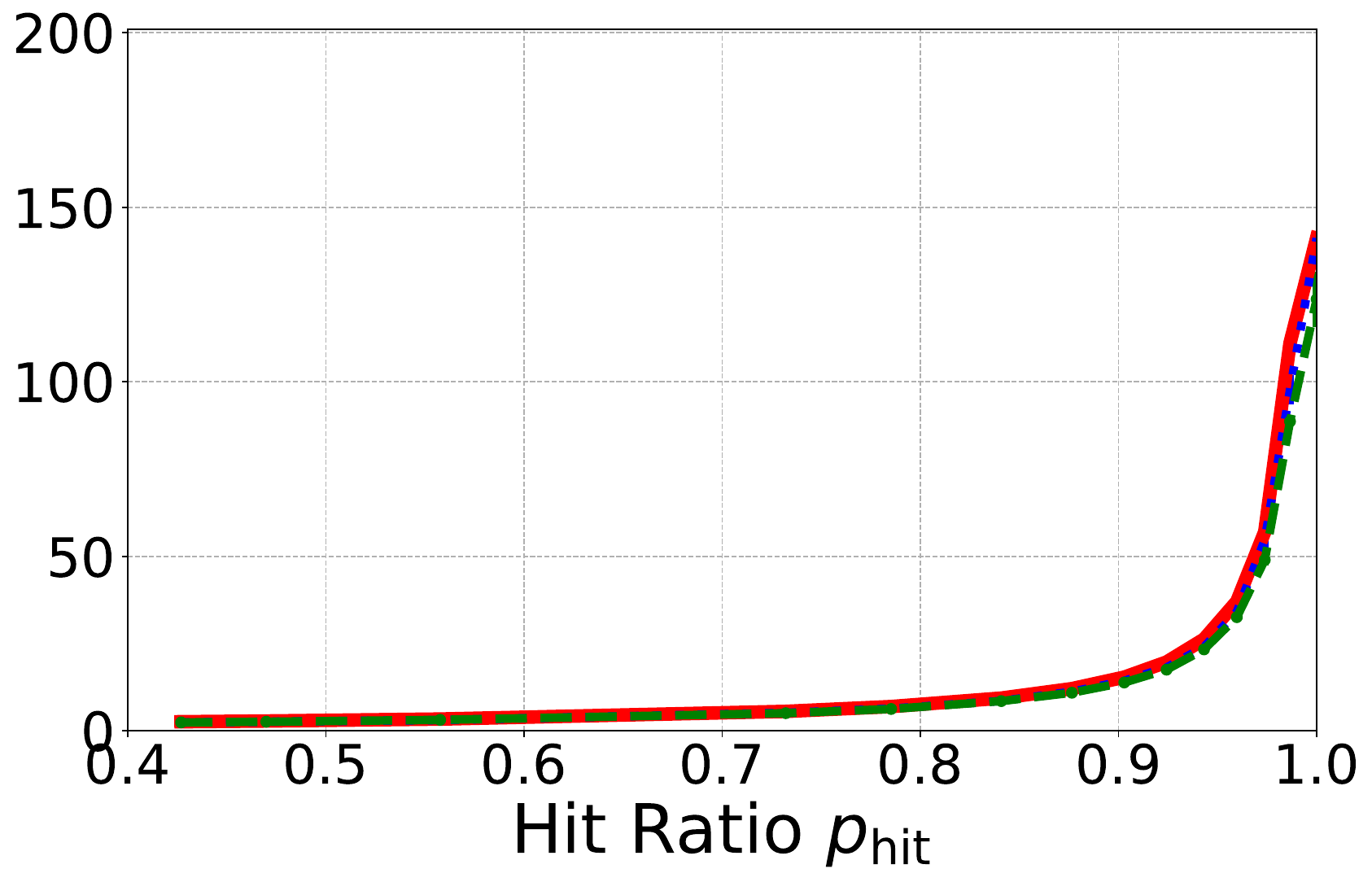}
  \caption{Disk latency 5 $\mu$s}
\end{subfigure}
\caption{Results for theory, implementation and simulation under a FIFO cache.   For all three curves, the throughput of the FIFO cache always increases at higher hit ratios under different disk latencies. }
\label{fig:FIFO-simu}
\end{figure*}

Equation (\ref{eqn:XFIFO100}) represents an {\em upper bound on throughput}, shown in red Figure~\ref{fig:FIFO-simu}(b).  
Recall that $0 < \E{S_{tail}} < 0.73$.  In the above analysis, we assumed that $\E{S_{tail}} = 0$ because we wanted an upper bound on $X$.  If instead, we had used any value of $\E{S_{tail}}$ in the range $(0, 0.73)$, $X$ would only change by $<0.5\%$.  

What's interesting about (\ref{eqn:XFIFO100}) is that both terms in the bound of $X$ increase with $p_{hit}$.   This contrasts with the expressions of throughput for LRU, where increasing $p_{hit}$ decreased the second term in the bound of $X$.

The above analysis assumed that $\E{Z_{disk}}= 100 \mu s$.  We can likewise redo the analysis for the case where $\E{Z_{disk}} = 5 \mu s$, obtaining:  

\begin{eqnarray}
X_{\mbox{\tiny{FIFO}}} \leq  \min\left( \frac{72}{6.24  - 5.73 p_{hit} } \ , \ \frac{1}{0.73 - 0.73 p_{hit}} \right)
\label{eqn:XFIFO5}
\end{eqnarray}

Likewise, when $\E{Z_{disk}} = 500 \mu s$, we obtain:

\begin{eqnarray}
X_{\mbox{\tiny{FIFO}}}  \leq  \min\left( \frac{72}{501.24  - 500.73 p_{hit} } \ , \ \frac{1}{0.73 - 0.73 p_{hit}} \right)
\label{eqn:XFIFO500}
\end{eqnarray}

The bounds in (\ref{eqn:XFIFO100}), (\ref{eqn:XFIFO5}), and (\ref{eqn:XFIFO500}) are shown via the red solid lines in Figure~\ref{fig:FIFO-simu}.

\noindent\textbf{Results of the three-pronged approach.}
Our analysis for FIFO shows that increasing the hit ratio {\em always} leads to higher throughput, regardless of the mean disk latency.  The queueing theory shows this mathematically.  More intuitively, the bottleneck device (the head update) is now always in the {\em miss} path, not in the hit path. Therefore, increasing the hit ratio does not result in increased demand on the bottleneck device, and hence does not lead to deleterious effects on throughput.  
Our simulation of the queueing network follows the same process as Section~\ref{ss:simulation}, and our implementation follows the process of Section~\ref{ss:implementation}. Results of simulation are shown in blue dotted lines and results of implementation are shown via green dashed lines in Figure~\ref{fig:FIFO-simu}. These agree within 5\%.

\subsection{Probabilistic LRU}


Under Probabilistic LRU there is an additional parameter $q$ that controls how close the algorithm is to LRU (lower $q$) versus FIFO (higher $q$).   
As always, there is a global linked list; however, now the ordering of items in the linked list is a mixture of FIFO and LRU.


\begin{figure*}[ht]
\begin{center}
\begin{subfigure}{0.7\linewidth}
    \includegraphics[width=\linewidth]{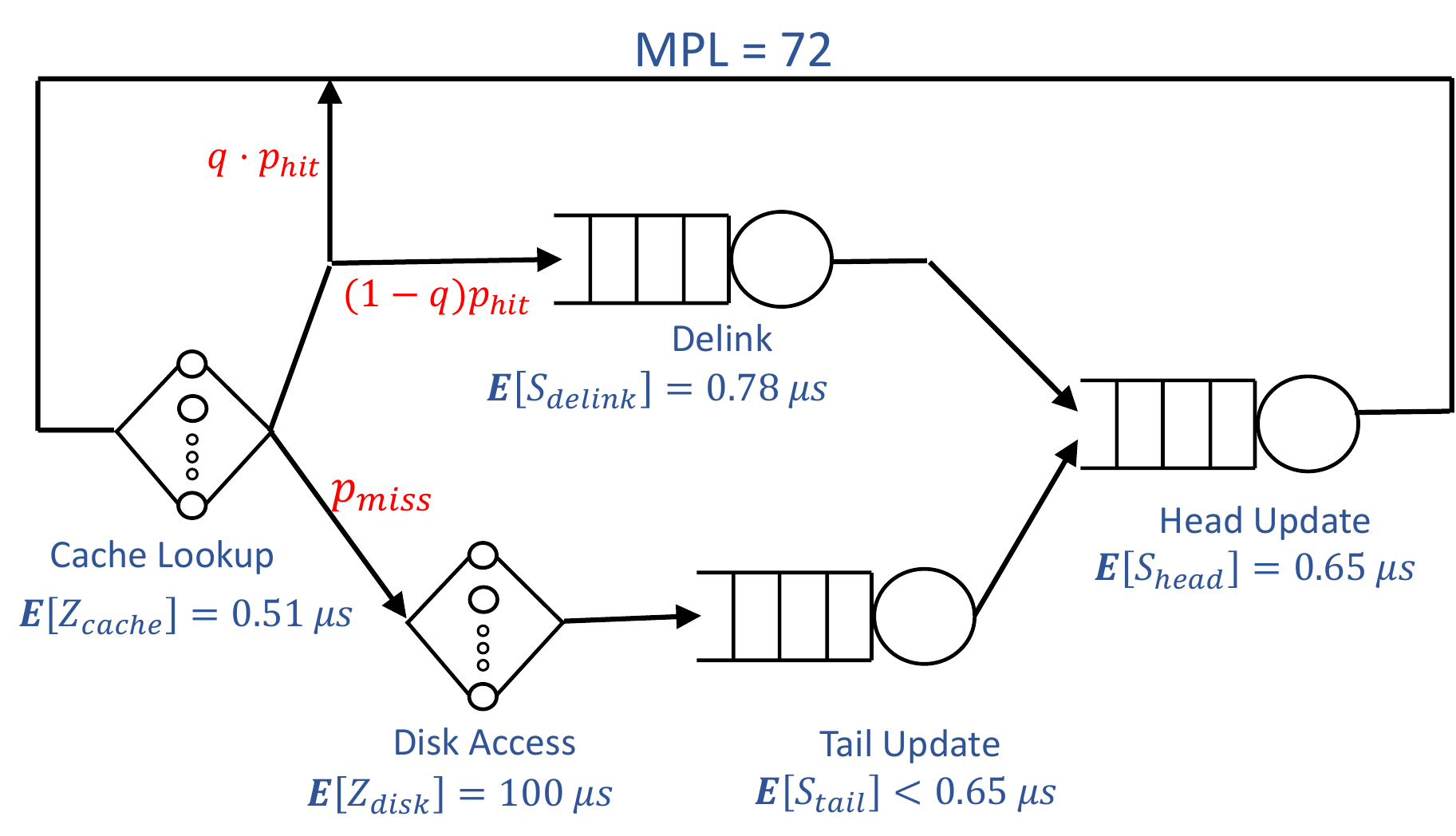}
    \caption{Queueing model of Probabilistic LRU cache with $q=0.5$.}
\end{subfigure}
\quad
\begin{subfigure}{0.7\linewidth}
    \includegraphics[width=\linewidth]{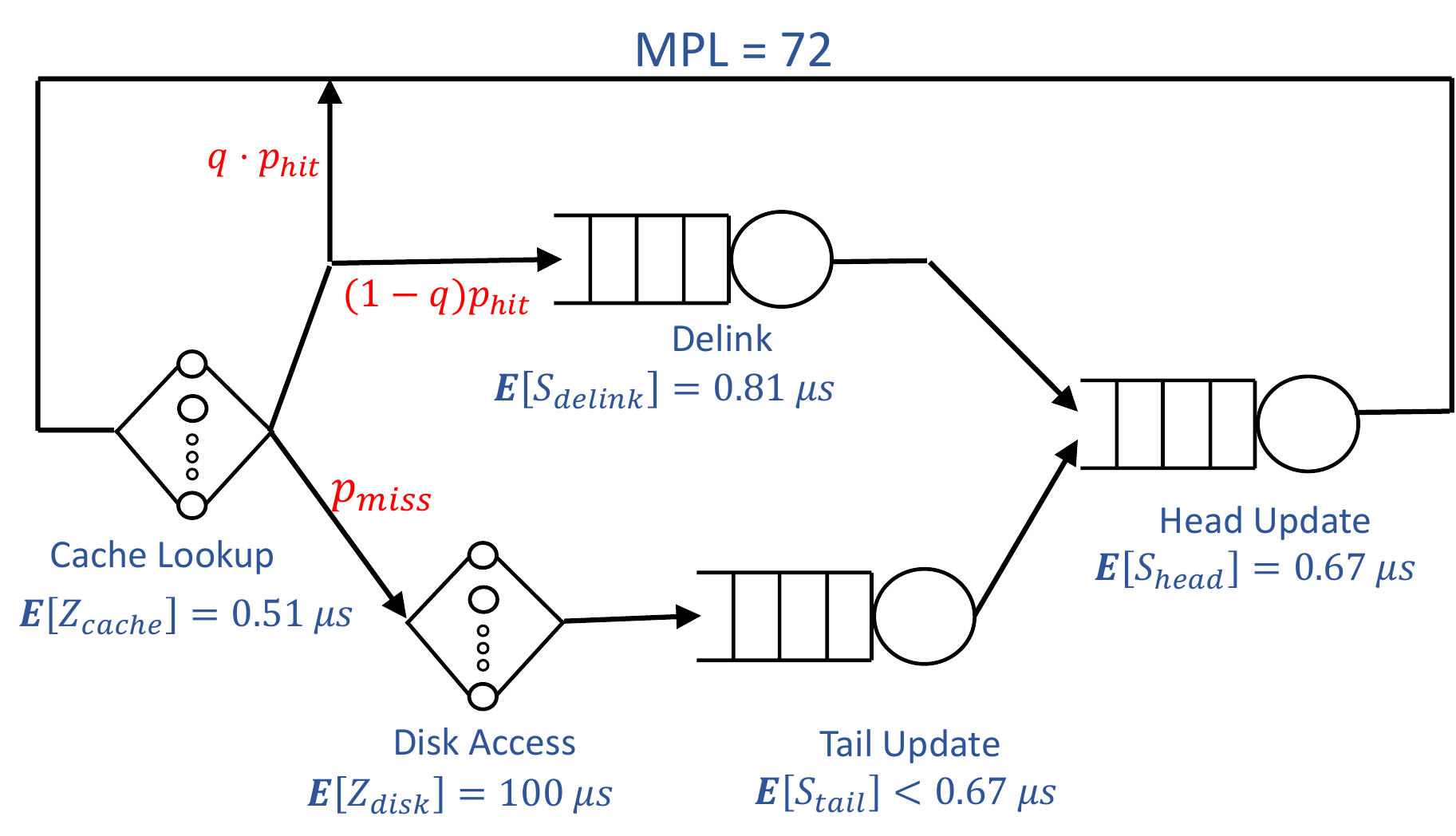}
    \caption{Queueing model of Probabilistic LRU cache with $q=1-\frac{1}{72}$. Notice that the numbers change slightly.}
\end{subfigure}
\end{center}
\vspace{-1.6em}
\caption{Queueing model of Probalistic LRU cache.}
\label{fig:ProbLRUcache}
\end{figure*}


If a request for item $d$ is a hit then, with probability $q$, {\em nothing} happens to the global linked list (as in FIFO), and, with probability $1-q$, we follow LRU.

If the request is a miss, then three things need to happen:
\begin{enumerate}
\item The item $d$ needs to be read from disk.   
\item The item at the tail of the global linked list must be removed (a tail update). 
\item Item $d$ must be attached to the global list head (a head update operation).
\end{enumerate}

The service times for the usual operations turn out to be slightly affected by the value of $q$.  This is because $q$ affects the queue lengths, hence affecting the communication overhead, as explained in Sections~\ref{sss:model-ES}.
The queueing network for Probabilistic LRU is shown in the case of $q = 0.5$ (Figure~\ref{fig:ProbLRUcache}(a)) and $q = 1 - \frac{1}{72}$ (Figure~\ref{fig:ProbLRUcache}(b)).   We chose these values because it turns out that $q$ has to be extremely high, $\geq 1 - \frac{1}{N}$, to show FIFO-like behavior.  For all other values of $q$, our analysis (and implementation), shows LRU-like behavior.  This is an interesting finding in its own right.


{\color{chighlight}
We follow the usual analysis approach
in Figure~\ref{fig:ProbLRUcache}.
For both networks, the {\em mean think time} of the system is: 
\begin{eqnarray*}
    \E{Z} & = & \E{Z_{cache}} + p_{miss} \cdot \E{Z_{disk}} 
     =  100.51 - 100 p_{hit}
\end{eqnarray*}

For the network in Figure~\ref{fig:ProbLRUcache}(a), we end up with the following results: 

\begin{eqnarray*}
D_{Delink} & = & 0.5 p_{hit}  \cdot 0.78 \\
D_{tail} & < & (1 - p_{hit})  \cdot 0.65 \\
D_{head} & = & (0.5 p_{hit} + 1 - p_{hit}) \cdot 0.65 
 = (1 - 0.5 p_{hit}) \cdot 0.65 \\
D_{max} & = &  \max\left(0.39 p_{hit}, 0.65 - 0.325 p_{hit}\right) \\
D &=& D_{Delink} + D_{tail} + D_{head}
\end{eqnarray*}

Because $0 < D_{tail} < (1 - p_{hit})  \cdot 0.65$, we have upper and lower bounds on $D$ as follows:
\begin{eqnarray*}
  0.65 + 0.065 p_{hit} < & D & < 0.65 + 0.065 p_{hit} + (1 - p_{hit}) \cdot 0.65 \\
0.65 + 0.065 p_{hit} <  &D & < 1.3 - 0.585 p_{hit}  
\end{eqnarray*}

Substituting the above expressions into \cite[Theorem 7.1]{PerformanceModeling13}, we find  that when $q = 0.5$:

\vspace{-0.8em}
\begingroup
\everymath{\scriptstyle}
\scriptsize
\begin{eqnarray*}
X_{\tiny \mbox{Prob-LRU}(q=0.5)}^{\E{Z_{disk}} = 100 \mu s}
\leq  
\min\left( \frac{72}{101.16  - 99.94 p_{hit} } \ , \ \frac{1}{\max(0.39 p_{hit}, 0.65 - 0.325 p_{hit})} \right) 
\label{eqn:XProbLRU2_100} \\
X_{\tiny \mbox{Prob-LRU}(q=0.5)}^{\E{Z_{disk}} = 5 \mu s} 
\leq  
\min\left( \frac{72}{6.16  - 4.94 p_{hit} } \ , \ \frac{1}{\max(0.39 p_{hit}, 0.65 - 0.325 p_{hit})} \right) 
\label{eqn:XProbLRU2_5} \\
X_{\tiny \mbox{Prob-LRU}(q=0.5)}^{\E{Z_{disk}} = 500 \mu s} 
\leq  
\min\left( \frac{72}{501.16  - 499.94 p_{hit} } \ , \ \frac{1}{\max(0.39 p_{hit}, 0.65 - 0.325 p_{hit})} \right) 
\label{eqn:XProbLRU2_500} 
\end{eqnarray*}
\endgroup

Likewise, we get the following upper bounds when $q = 1 - \frac{1}{72} = 0.986$:

\vspace{-0.8em}
\begingroup
\everymath{\scriptstyle}
\scriptsize
\begin{eqnarray*}
X_{\tiny \mbox{Prob-LRU}(q=0.986)}^{\E{Z_{disk}} = 100 \mu s} 
\leq  
\min\left( \frac{72}{101.18  - 100.65 p_{hit} } \ , \ \frac{1}{\max(0.011 p_{hit}, 0.67 - 0.656 p_{hit})} \right) 
\label{eqn:XProbLRU72_100} \\
X_{\tiny \mbox{Prob-LRU}(q=0.986)}^{\E{Z_{disk}} = 5 \mu s} 
\leq  
\min\left( \frac{72}{6.18  - 5.65 p_{hit} } \ , \ \frac{1}{\max(0.011 p_{hit}, 0.67 - 0.656 p_{hit})} \right) 
\label{eqn:XProbLRU72_5} \\
X_{\tiny \mbox{Prob-LRU}(q=0.986)}^{\E{Z_{disk}} = 500 \mu s} 
\leq  
\min\left( \frac{72}{501.18  - 500.65 p_{hit} } \ , \ \frac{1}{\max(0.011 p_{hit}, 0.67 - 0.656 p_{hit})} \right) 
\label{eqn:XProbLRU72_500} 
\end{eqnarray*}
\endgroup

Note that we used the lower bound on $D$ to upper-bound the throughput.  However, as in our prior analyses, using the upper bound on $D$ will only change the results by $< 0.5\%$.
}

\begin{figure*}[t]
\centering
\begin{subfigure}{.33\textwidth}
  \centering
  \includegraphics[height=1.in]{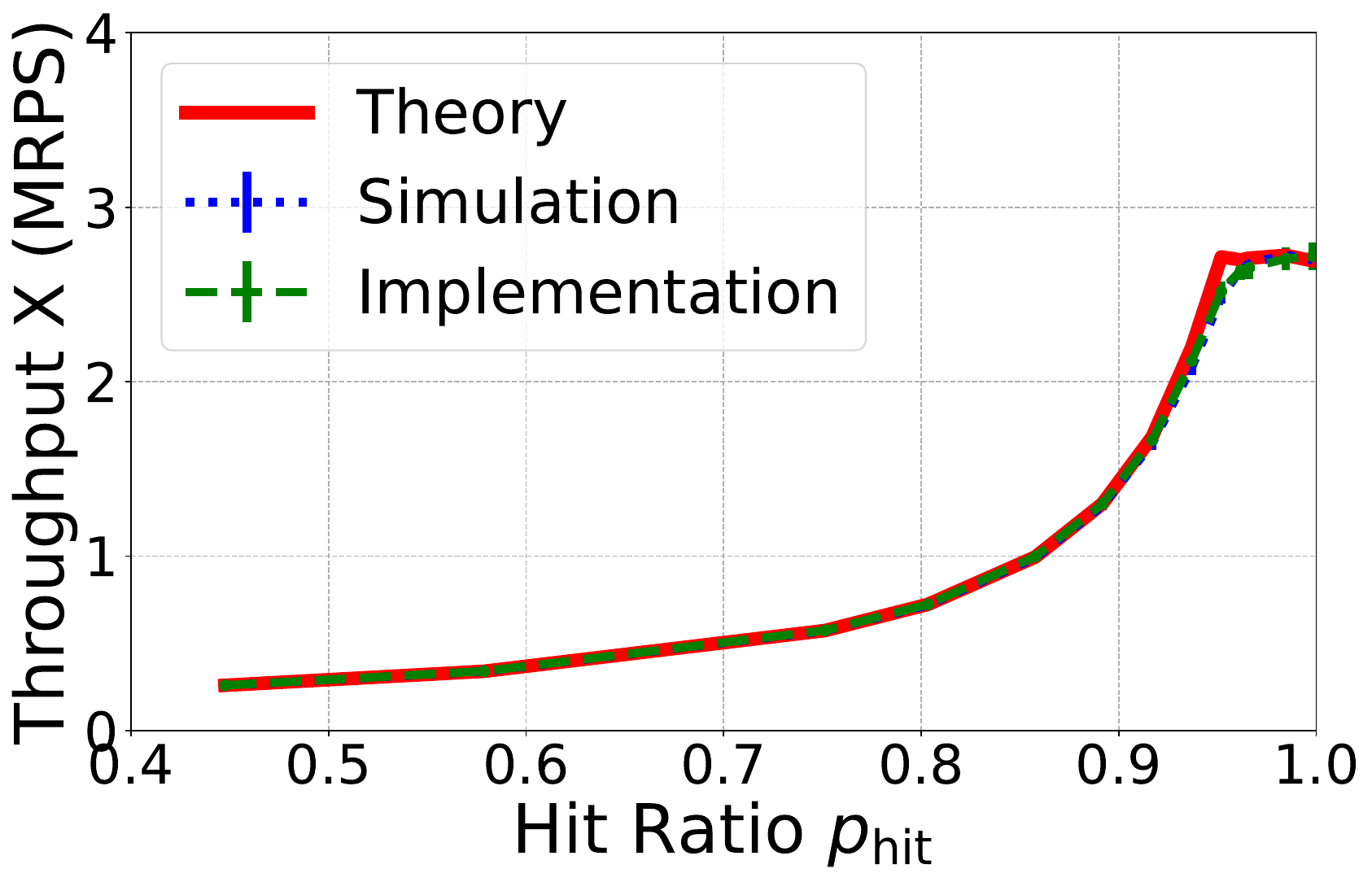}
  \caption{Disk latency 500 $\mu$s}
\end{subfigure}%
\begin{subfigure}{.33\textwidth}
  \centering
  \includegraphics[height=1.in]{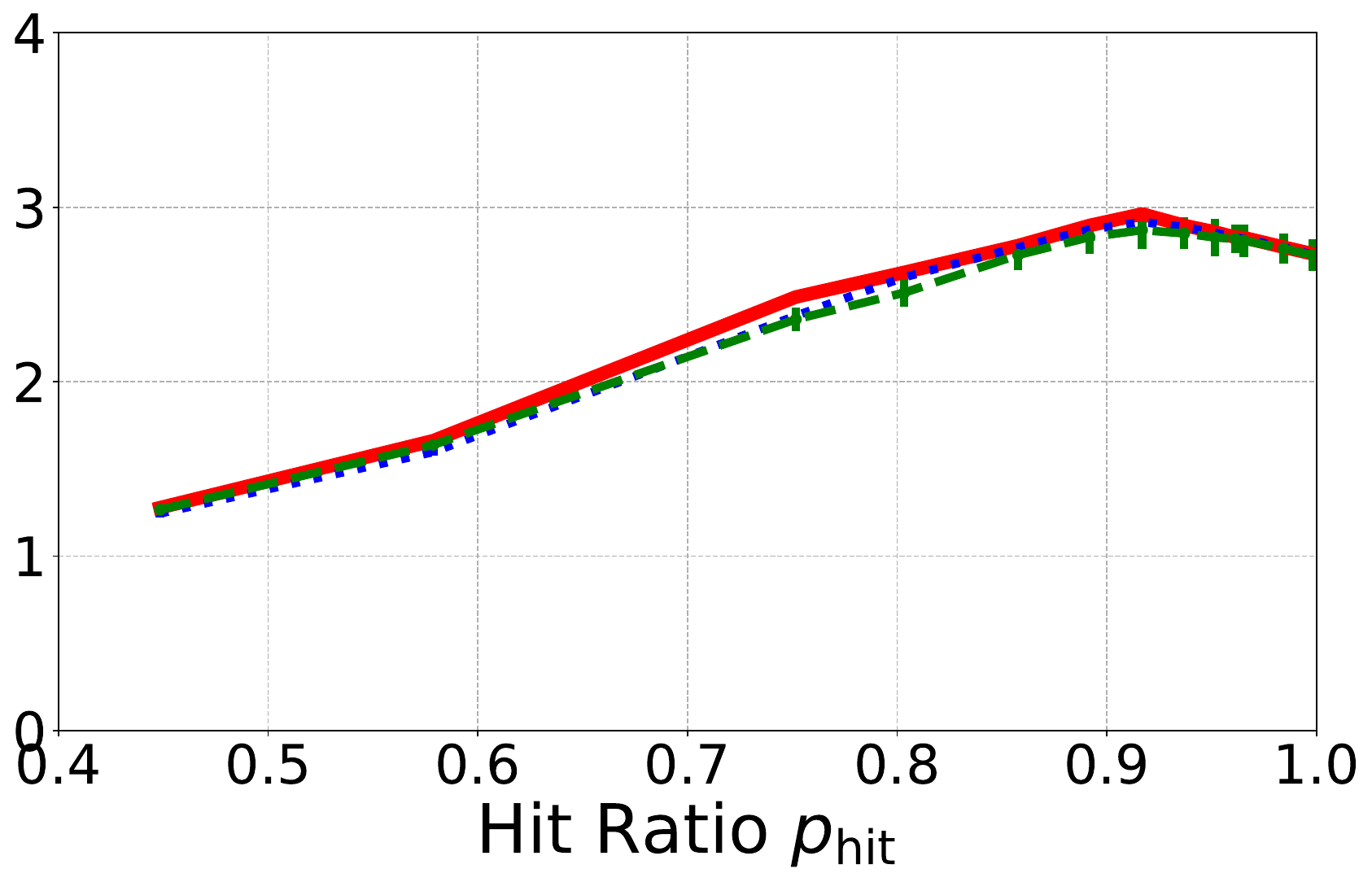}
  \caption{Disk latency 100 $\mu$s}
\end{subfigure}
\begin{subfigure}{.33\textwidth}
  \centering
  \includegraphics[height=1.in]{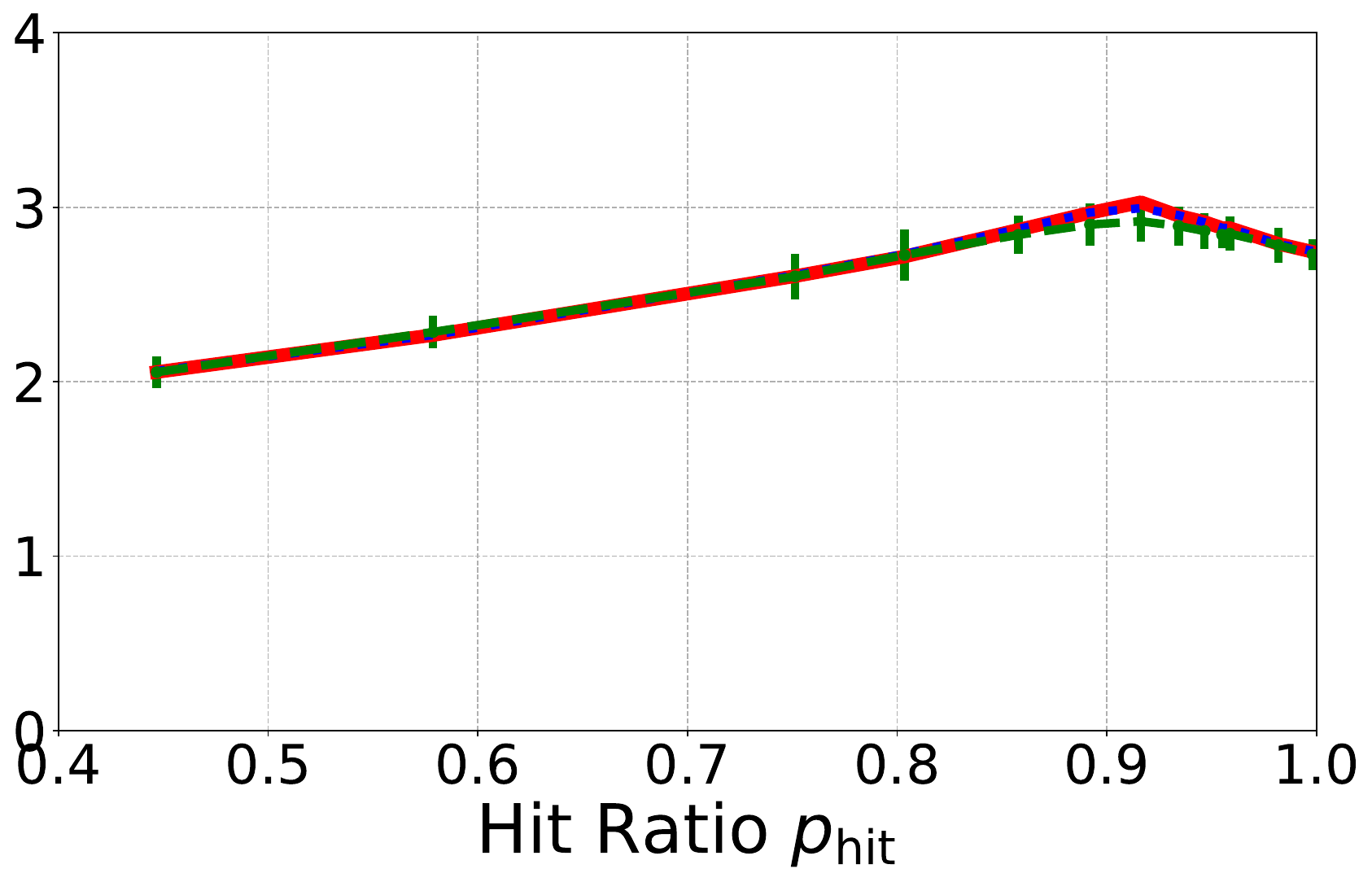}
  \caption{Disk latency 5 $\mu$s}
\end{subfigure}
\vspace{-0.8em}
\caption{Results for theory, implementation and simulation under a Probabilistic LRU cache with $q = 0.5$.   Throughput decreases at higher hit ratios under all disk latencies.}
\label{fig:probLRU_2-simu}
\end{figure*}

\begin{figure*}[t]
\centering
\begin{subfigure}{.33\textwidth}
  \centering
  \includegraphics[height=1.in]{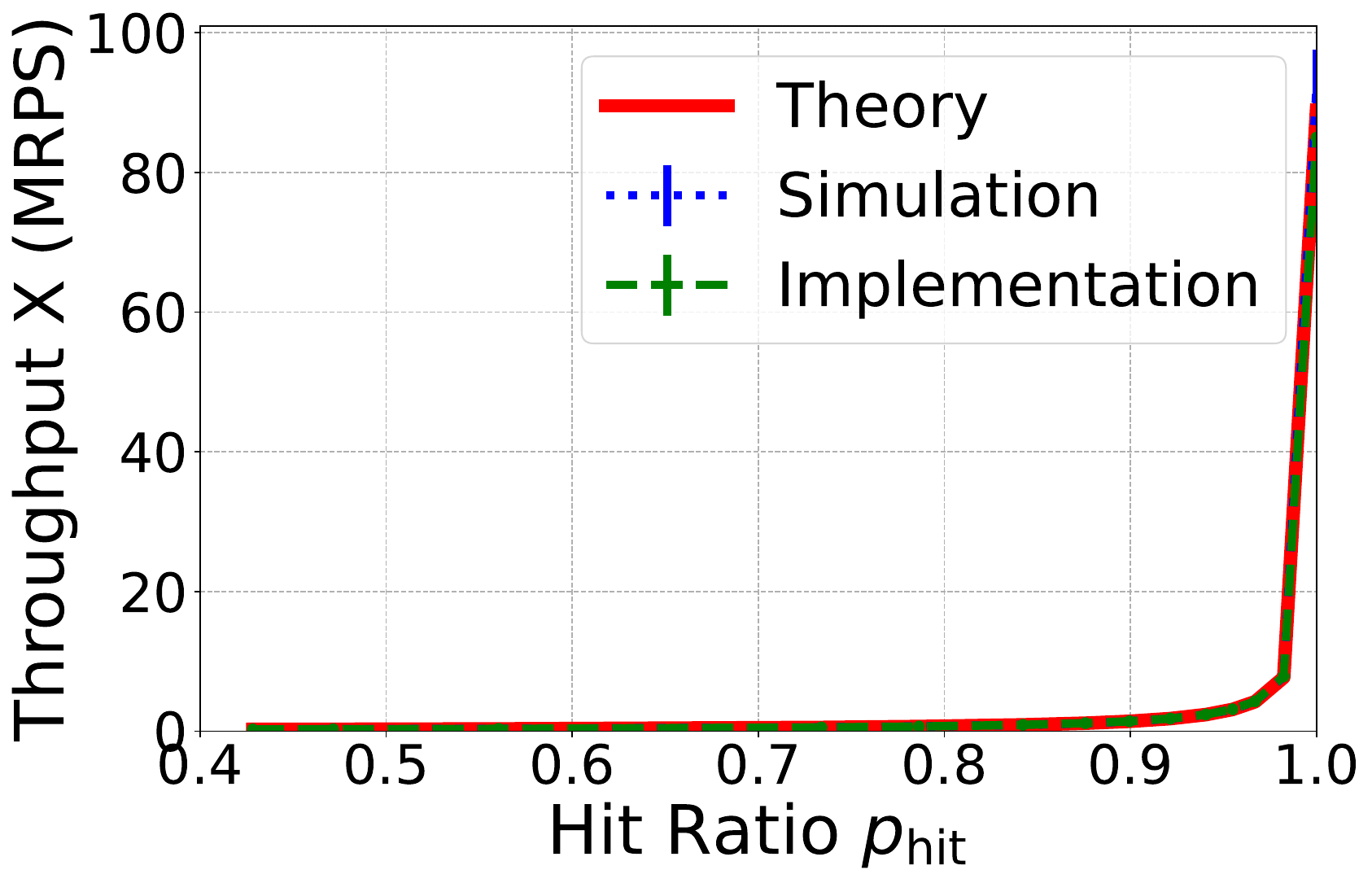}
  \caption{Disk latency 500 $\mu$s}
\end{subfigure}%
\begin{subfigure}{.33\textwidth}
  \centering
  \includegraphics[height=1.in]{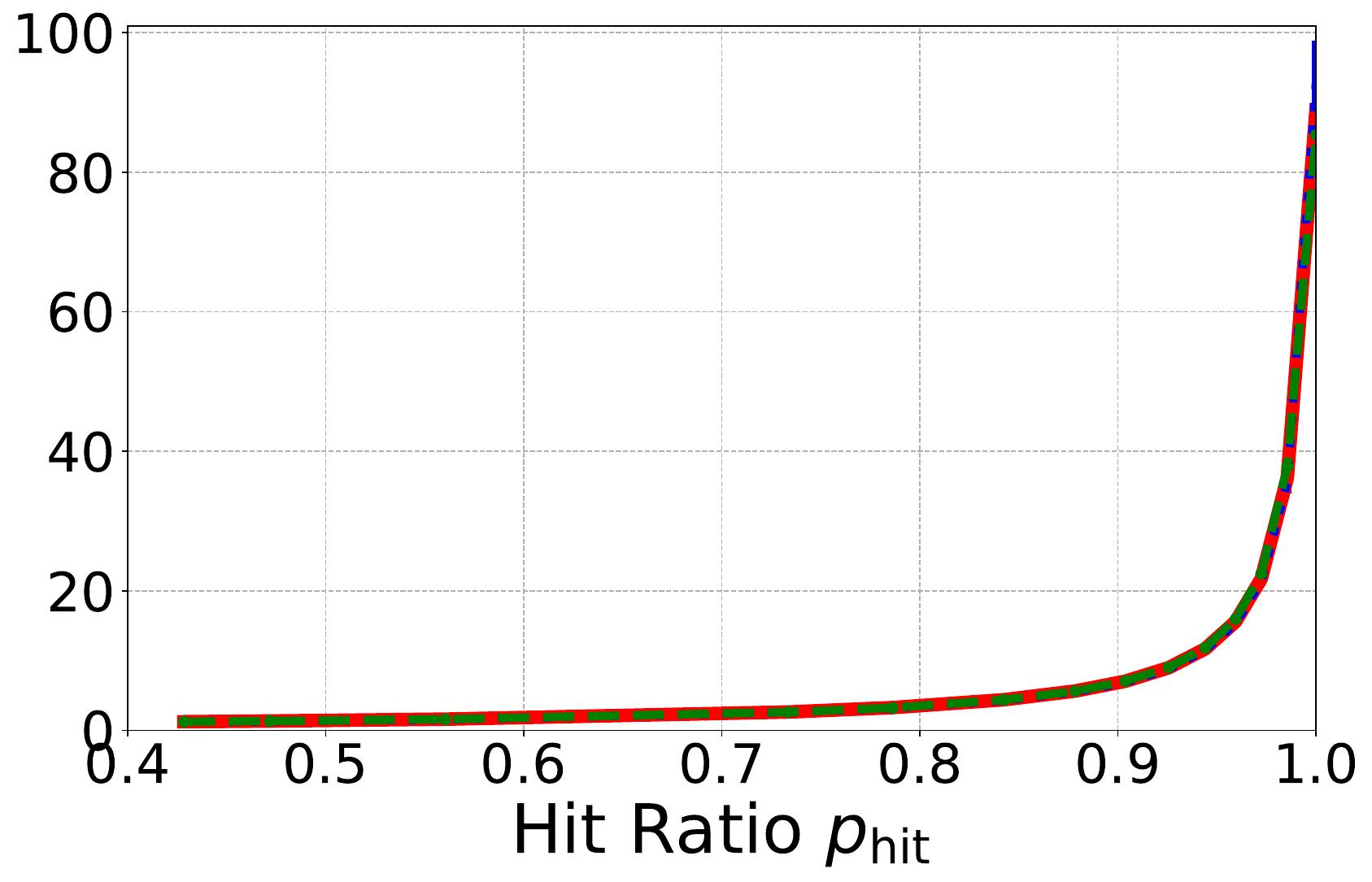}
  \caption{Disk latency 100 $\mu$s}
\end{subfigure}
\begin{subfigure}{.33\textwidth}
  \centering
  \includegraphics[height=1.in]{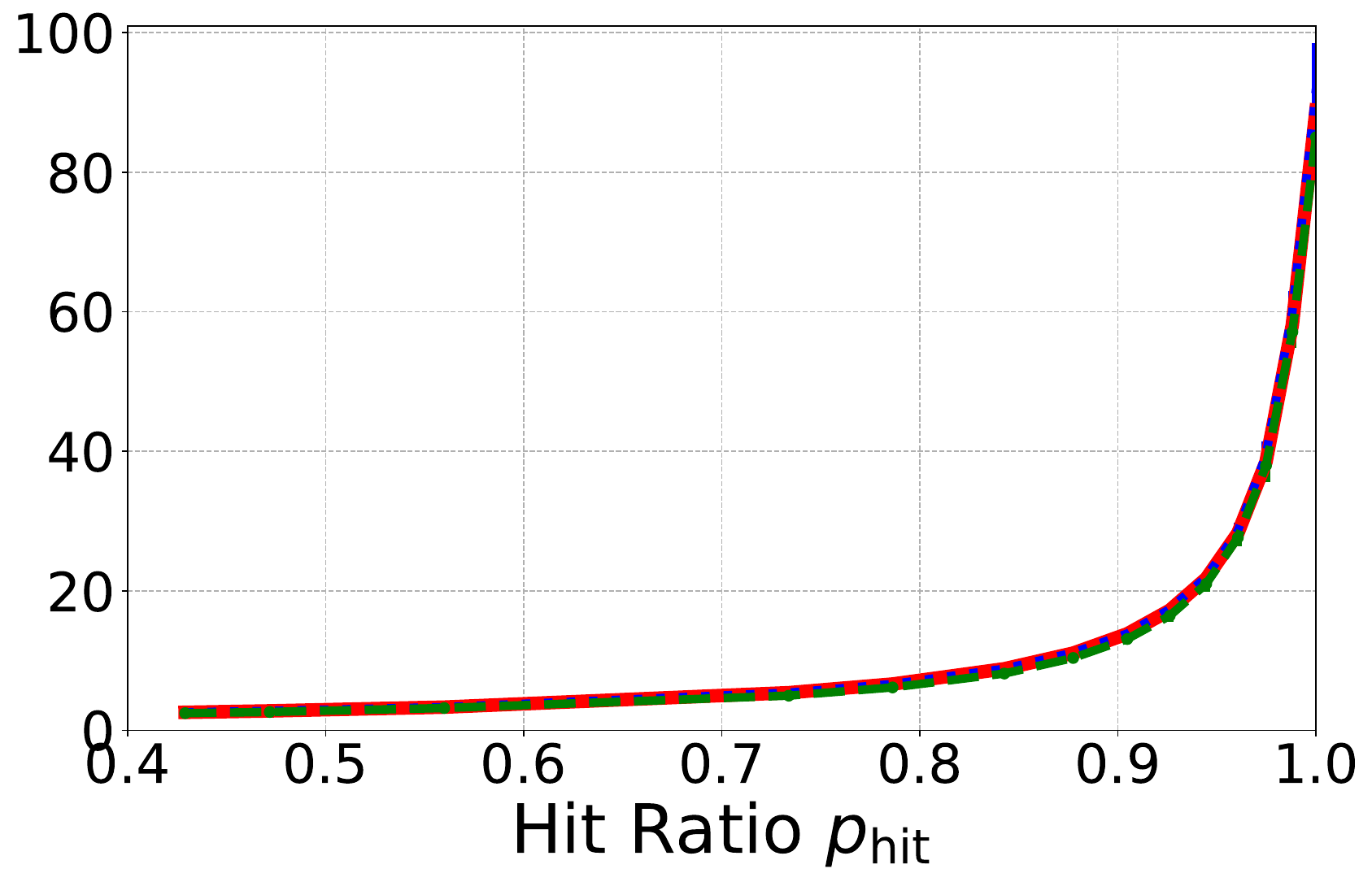}
  \caption{Disk latency 5 $\mu$s}
\end{subfigure}
\vspace{-0.8em}
\caption{Results for theory, implementation and simulation under Probabilistic LRU with $q = 1 - \frac{1}{72} = 0.986$.   For all three curves, throughput increases with hit ratio. \vspace{-1.6em}}
\label{fig:probLRU_72-simu}
\end{figure*}

\noindent\textbf{Results of the three-pronged approach.}
Figures~\ref{fig:probLRU_2-simu} and~\ref{fig:probLRU_72-simu} show the results of analysis (red solid lines), simulation (blue dotted lines) and implementation (green dashed lines). Again simulation and implementation agree within 5\% and the analytic upper bound provides an excellent indication of the trends.

The behavior of Probabilistic LRU is highly dependent on the $q$ parameter.  When {\em $q$ is not very high}, we see that the throughput starts decreasing beyond hit ratio $p^*_{hit}$.  The queueing theory shows this fact mathematically.  More intuitively, given that $q$ is not high, many requests need to go through the delink queue, which becomes the bottleneck, resulting in higher queueing times and reduced throughput.

By contrast, when  {\em $q$ is very high}, we basically skip the delink operation.  Hence our network behaves very similarly to FIFO.  Now the bottleneck device is on the {\em miss} path, and hence is not affected by increasing the hit ratio.

\subsection{CLOCK}

\begin{figure}[t]
\centerline{\psfig{file=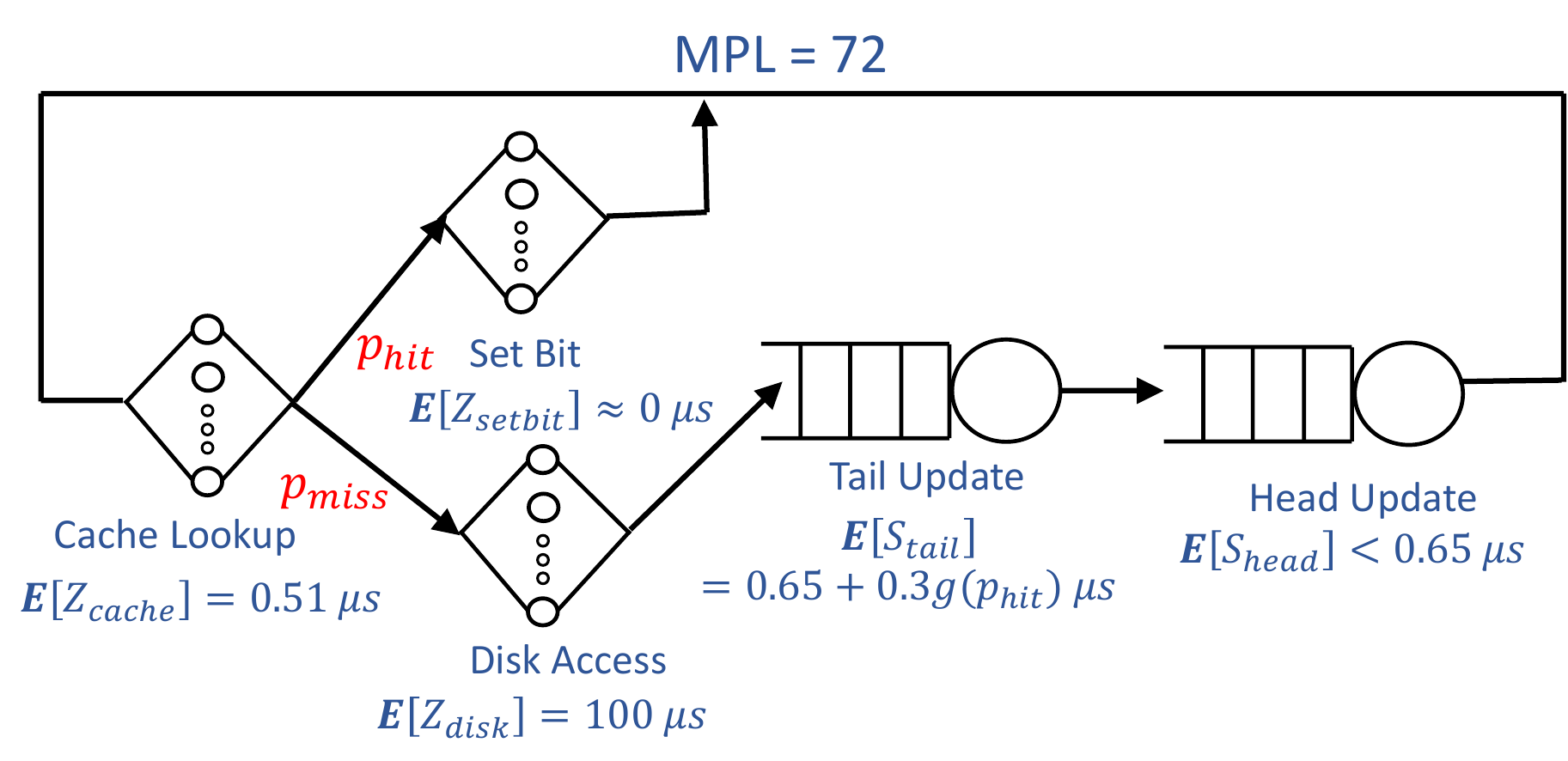,height=1.6in}}
\caption{Queueing model of CLOCK cache. \vspace{-0.8em}}
\label{fig:CLOCKcache}
\end{figure}

The CLOCK cache eviction policy operates a global linked list, ordered like FIFO, however every object that is accessed gets a ``second chance" to live before eviction.   As in FIFO, each object $d$ is appended to the head of the linked list and moves down towards the tail of the list as new objects are appended to the head. Each object is given a special bit set to $0$.  If object $d$ gets to the tail of the list, and its bit is still $0$, then it is removed from the tail.  However, if $d$ is accessed before it gets to the tail, then $d$'s bit is set to $1$.  Note that $d$'s position in the list does not change. When $d$ gets to the tail of the list, because its bit is $1$, it is skipped over for eviction, provided that there are other candidates with a $0$ bit. The CLOCK queueing network is shown in Figure~\ref{fig:CLOCKcache}.

\begin{figure*}[t]
\centering
\begin{subfigure}{.33\textwidth}
  \centering
  \includegraphics[height=1.in]{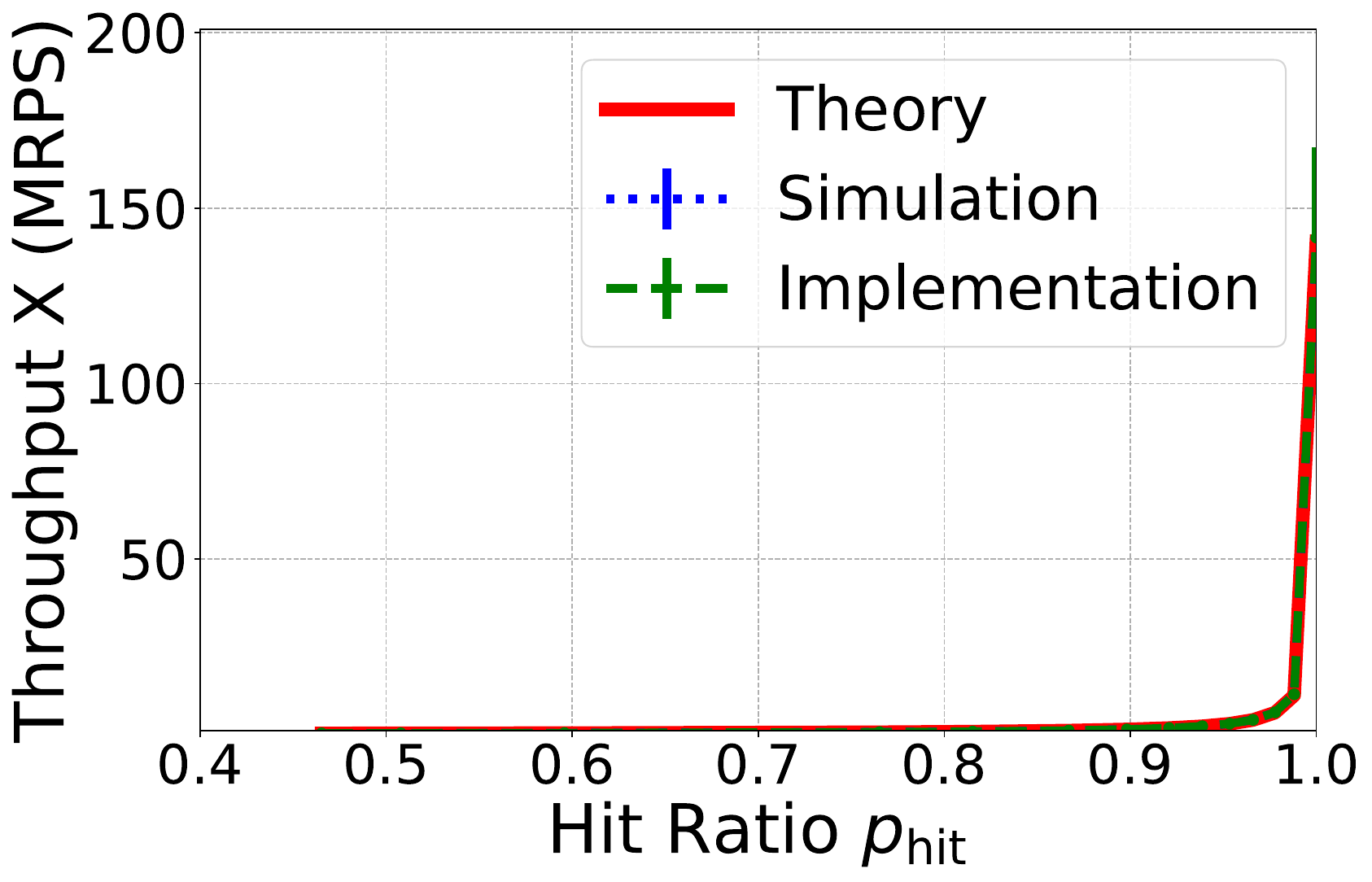}
  \caption{Disk latency 500 $\mu$s}
\end{subfigure}%
\begin{subfigure}{.33\textwidth}
  \centering
  \includegraphics[height=1.in]{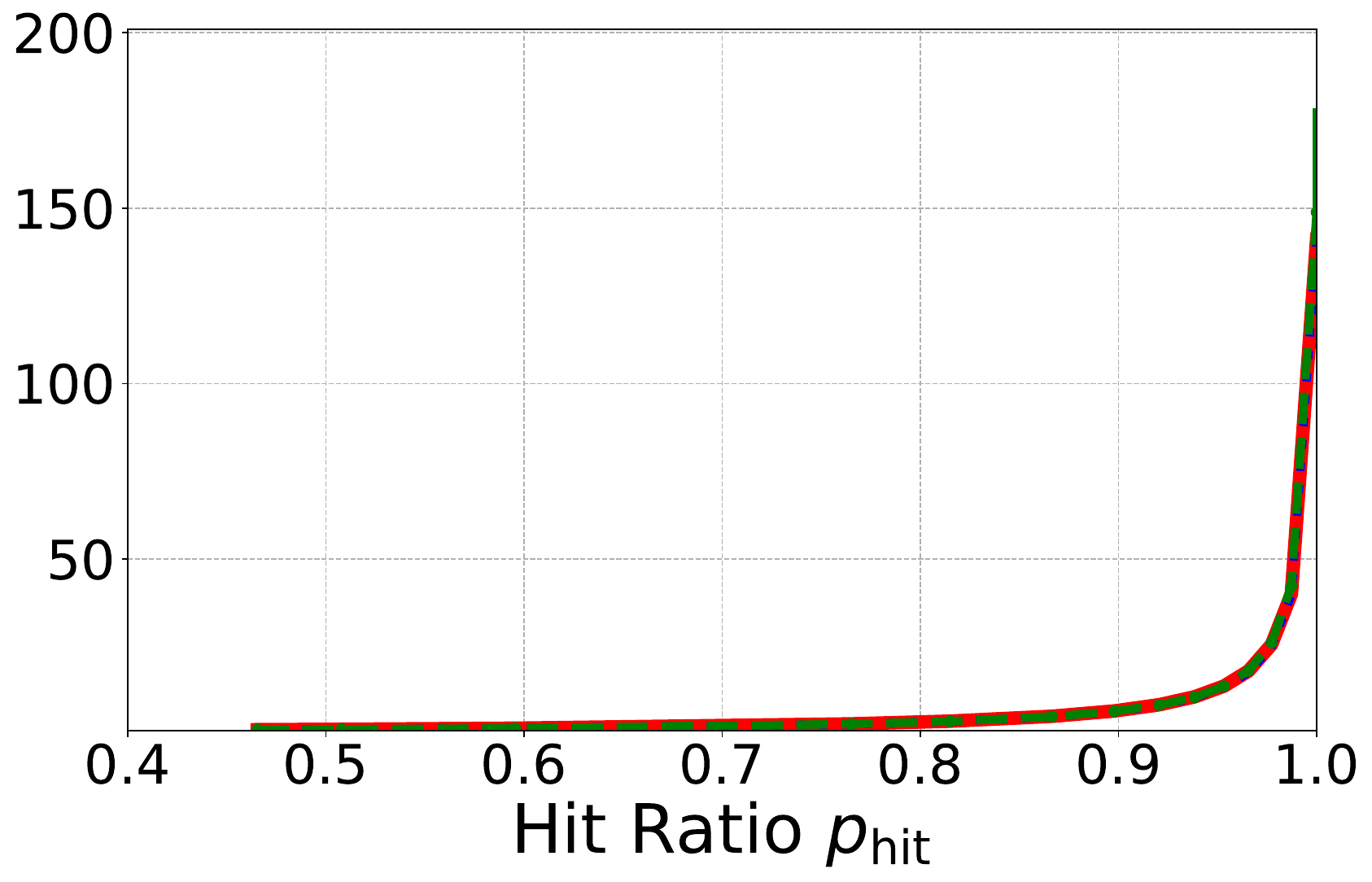}
  \caption{Disk latency 100 $\mu$s}
\end{subfigure}
\begin{subfigure}{.33\textwidth}
  \centering
  \includegraphics[height=1.in]{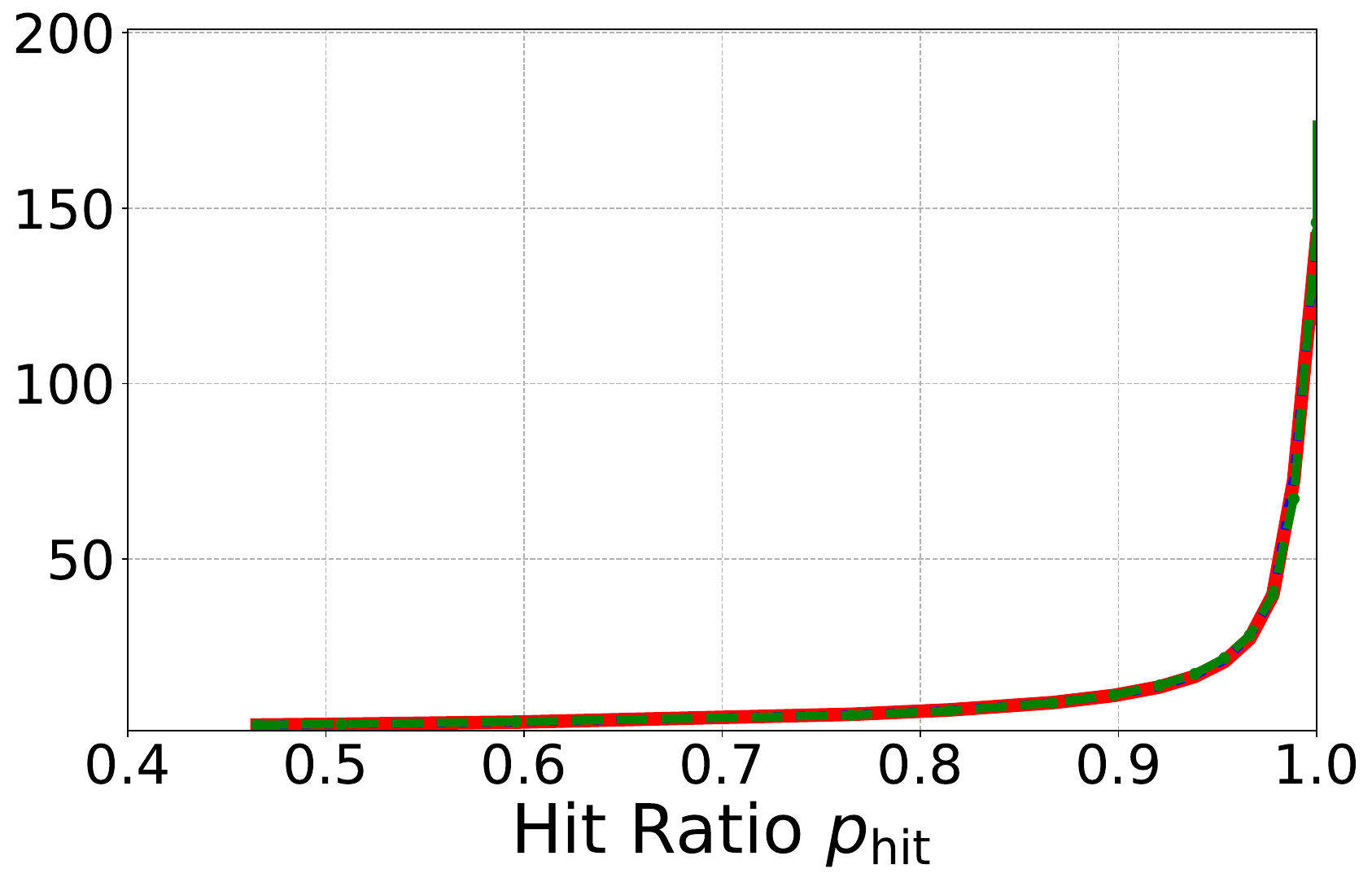}
  \caption{Disk latency 5 $\mu$s}
\end{subfigure}
\caption{Results for theory, implementation and simulation under a CLOCK cache.   For all three curves, the throughput of the CLOCK cache always increases with hit ratio under different disk latencies. }
\vspace{-1.2em}
\label{fig:CLOCK-simu}
\end{figure*}


{\color{chighlight}
Our specific implementation works as follows:  Let $n$ be the length of the linked list. When we're looking for an item to evict from the tail of the list, we start with the $n$th item (the last one) and check its bit.  If it is $0$, we evict it.  If not, then we move to the $(n-1)$th item.  If that is $0$, we evict it.  If not, then we move to the $(n-2)$th item.  If that is $0$, we evict it.  If not, then we remove the $(n-3)$th item, regardless of its bit.  
Thus the time, $S_{tail}$, needed for a tail update depends on how long it takes to find a $0$ bit.  If $p_{hit}$ is high, the time is longer, since more items have a $1$ bit.  We have carefully modeled $\E{S_{tail}}$, and we find that $\E{S_{tail}}$ has two components, representing (i) the time to read the tail bit of the $n$th item and do a regular tail update and (ii) the additional time needed to potentially read the $(n-1)$th bit and $(n-2)$th bit and so on.   
Our measurements show that $$\E{S_{tail}} = 0.65 + 0.3 \cdot g(p_{hit}),$$
where 
\begin{eqnarray*}
g(x) & = & 2.43 \cdot 10^{-5} e^{11.24x} + 0.187. \label{eqn:gfunc}
\end{eqnarray*}

The queueing model for a CLOCK cache is shown in Figure~\ref{fig:CLOCKcache}.  As expected, this basically looks like a FIFO queueing network.  The biggest difference is that the tail update queue is now the bottleneck.  
The other difference is that a hit in the cache requires setting the bit of the hit item to $1$ instead of $0$.  It turns out that setting this bit takes  $<0.01 \mu s$ time, so we consider that to be $0$.

\subsubsection*{Analysis of the queueing network model} Following the usual analysis approach, the {\em mean think time} of the system is: \begin{eqnarray*}
    \E{Z} & = & \E{Z_{cache}} + p_{miss} \cdot \E{Z_{disk}} 
     =  100.51 - 100 p_{hit}
\end{eqnarray*}

For each queue, we now compute the {\em device demand}:

\begin{eqnarray*}
D_{tail} & = & (1 - p_{hit})  \cdot (0.65 + 0.3 \cdot g(p_{hit})) \\
D_{head} & < & (1 - p_{hit}) \cdot 0.65 \\
D & = &  D_{tail} + D_{head}\\
D & > & (1 - p_{hit}) \cdot (0.65 + 0.3 \cdot g(p_{hit}))  \\
D & < & (1 - p_{hit}) \cdot (0.65 + 0.3 \cdot g(p_{hit})) + (1 - p_{hit}) \cdot 0.65
\end{eqnarray*}


The bottleneck device is the tail update, so:
  $$D_{max} = (1 - p_{hit})  \cdot (0.65 + 0.3 \cdot g(p_{hit})). $$
Substituting the above expressions into   \cite[Theorem 7.1]{PerformanceModeling13}, we obtain an upper bound on throughput in the case where $\E{Z_{disk}} = 100 \mu s$:



\begin{eqnarray*}
X_{\tiny \mbox{CLOCK}} & \leq  & \min\left(A, B\right) \\
A & = & \frac{72}{101.16 + 0.3 \cdot g(p_{hit}) - (100.65 + 0.3 \cdot g(p_{hit})) p_{hit}} \\
B & = & \frac{1}{(1 - p_{hit}) (0.65 + 0.3 \cdot g(p_{hit}))}
\end{eqnarray*}

\normalsize
We have derived similar bounds for the case where the disk latency is 500 $\mu s$ and 5 $\mu s$.  All our bounds are shown in Figure~\ref{fig:CLOCK-simu}. 

}

\noindent\textbf{Results of the three-pronged approach.}
Figure~\ref{fig:CLOCK-simu} shows the result of the theory, simulation, and implementation of CLOCK. 
We see
that increasing the hit ratio, $p_{hit}$, 
{\em always} leads to higher throughput, regardless of the mean disk latency.  The queueing theory shows this fact mathematically.  Intuitively, the bottleneck device (the tail update) is in the miss path, not in the hit path. Therefore, increasing the hit ratio this does not result in increased demand on the bottleneck device, and hence does not impinge on throughput.

\subsection{Segmented LRU}

The Segmented LRU (SLRU) policy is one of the more advanced policies that researchers and practitioners use~\cite{huang_analysis_2013}.  
The high-level idea in SLRU is that the global linked list of all objects in the cache is divided into two lists: the Probationary list ( denoted B for bottom) and the Protected list (T for top).


Objects initially enter the B list.
If an object is never accessed after being put on the B list, it will eventually leave the cache. 
However, if an object on the B List is accessed, 
then the object will be delinked from the B list and moved onto 
the T list.  The T list is an LRU list in the sense that it is sorted from most recently accessed to least recently accessed.   When an object on the T list is accessed, the object moves to the head of the T list.  When an object leaves the T list, it is moved back to the B list and the process repeats.  

\begin{figure}[t]
\centerline{\psfig{file=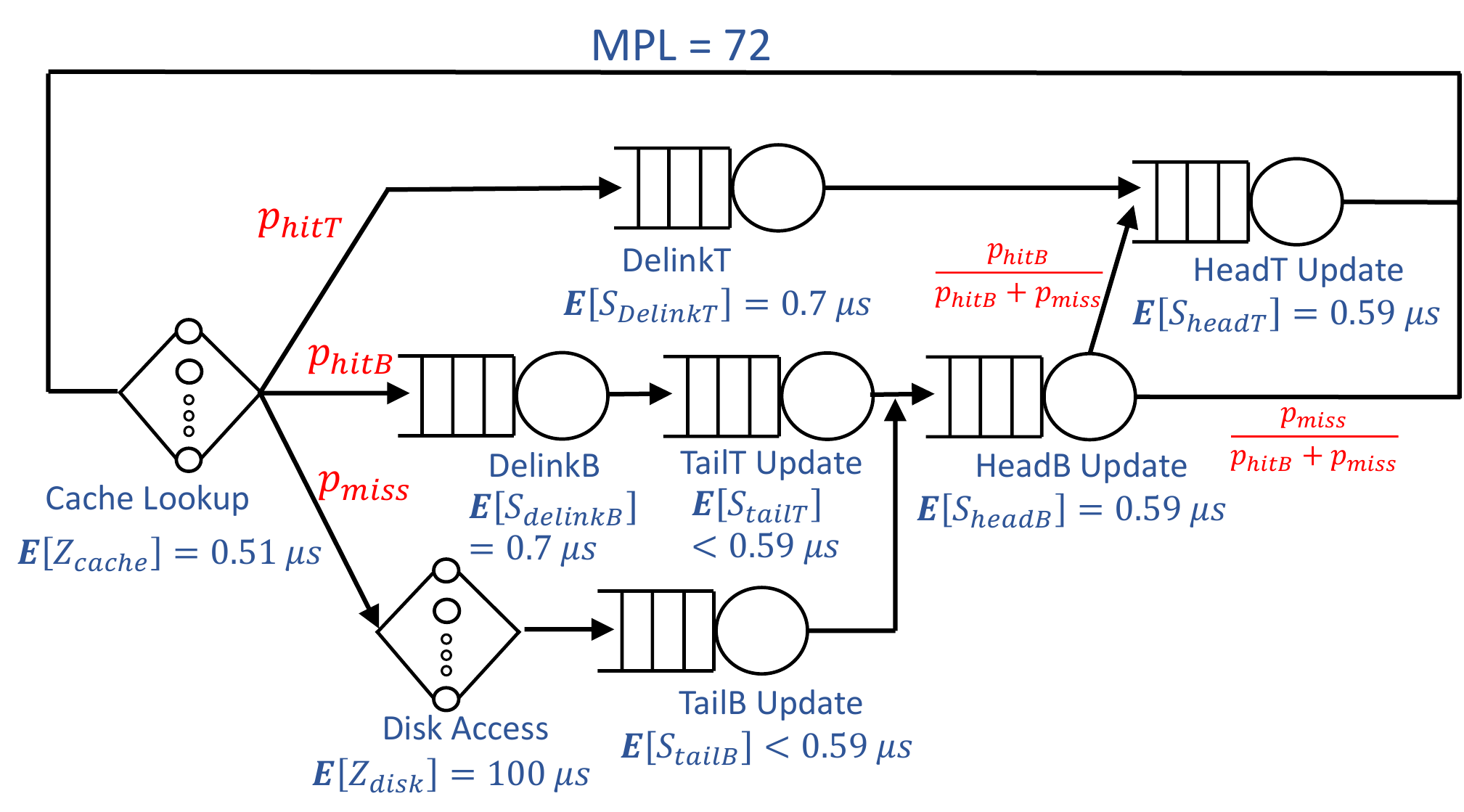,height=2.08in}}
\caption{Queueing model of Segmented LRU cache. }
\vspace{-1.6em}
\label{fig:SLRUcache}
\end{figure}

The queueing network for SLRU is shown in Figure~\ref{fig:SLRUcache}.  
When an object $d$ is first accessed, we do a cache lookup.  There are now three cases of what might happen.  If $d$ is currently in the T list (which runs LRU), then $d$ needs to be delinked from its current location and moved to the head of the B list (thus we have a delinkT operation followed a headT update).
If $d$ is currently in the B list, then $d$ is delinked from the B list and moved to the head of the T list.  When this happens, we need to remove the object that is at the tail of the T list.  That object is moved to the head of the B list.  
Finally, if $d$ is not in the cache (a ``miss"), then we need to find $d$ in the disk.  After finding $d$, we put it on the head of the B list and remove the object at the tail of the B list.  

To do a queueing analysis of the network in Figure~\ref{fig:SLRUcache}, we need two more things:
First, we need the service times of potential bottleneck operations. These are the same as the numbers in the LRU network.  Second, and more challenging, we need to understand the fraction of time that an object is found on the T list as opposed to the B list.  This is obviously a function of $p_{hit}$.  

{\color{chighlight}
We used our simulation to estimate this function and matched it to the following:
\begin{eqnarray*}
\P{\mbox{Object in T list} } & = & \ell(p_{hit}) \equiv -0.1144 \cdot p_{hit}^2 + 1.009 \cdot p_{hit} \\
\P{\mbox{Object in B list} } & = &  f(p_{hit}) \equiv p_{hit} - \ell(p_{hit}) 
\end{eqnarray*}

\subsubsection*{Analysis of the queueing network model}

We now analyze the closed queueing network in Figure~\ref{fig:SLRUcache}.  

The {\em mean think time} of the system is: 
\begin{eqnarray*}
    \E{Z} & = & \E{Z_{cache}} + p_{miss} \cdot \E{Z_{disk}} 
     =  100.51 - 100 p_{hit}
\end{eqnarray*}

For each queue, we now compute the {\em device demand}:

\vspace{-0.8em}
\begin{align*}
D_{DelinkT} &= \ell(p_{hit}) \cdot 0.7 & D_{tailT} &< f(p_{hit}) \cdot 0.59 \\
D_{DelinkB} &= f(p_{hit}) \cdot 0.7 & D_{tailB} &< (1 - p_{hit}) \cdot 0.59 \\
D_{headT} &= p_{hit} \cdot 0.59 & D_{headB} &= (1 - \ell(p_{hit})) \cdot 0.59
\end{align*}

where 
$$D_{max} =  \max\left(0.7 \cdot \ell(p_{hit}), 0.59 p_{hit}, 0.59 (1 - \ell(p_{hit}))\right).$$

The total demand, D, is the sum of all device demands:
\vspace{-0.2em}
$$D = D_{DelinkT} + D_{DelinkB} + D_{tailT} + D_{tailB} + D_{headT} + D_{headB}$$
\vspace{-0.8em}
\begin{eqnarray*}
D & > & 1.29 p_{hit} + 0.59 - 0.59 \cdot \ell(p_{hit}) \\
D & < & 1.29 p_{hit} + 1.18 - 1.18 \cdot \ell(p_{hit})
\end{eqnarray*}



Substituting the above expressions into \cite[Theorem 7.1]{PerformanceModeling13}, we obtain an upper bound on throughput in the case where 
$\E{Z_{disk}} = 100 \mu s$:

\vspace{-0.8em}
\begin{eqnarray*}
X_{\tiny \mbox{SLRU}} & \leq  & \min\left(A, B \right) \\
A & = & \frac{72}{101.1 - 88.71 p_{hit} - 0.59 \ell(p_{hit})} \\
B & = & \frac{1}{\max\left(0.7 \ell(p_{hit}), 0.59 (1 - \ell(p_{hit}))\right)}
\end{eqnarray*}

We have derived similar expressions for the case of $\E{Z_{disk}} = 500 \mu s$ and $\E{Z_{disk}} = 5 \mu s$.
}

\begin{figure*}[ht]
\centering
\begin{subfigure}{.32\textwidth}
  \centering
  \includegraphics[height=1.1in]{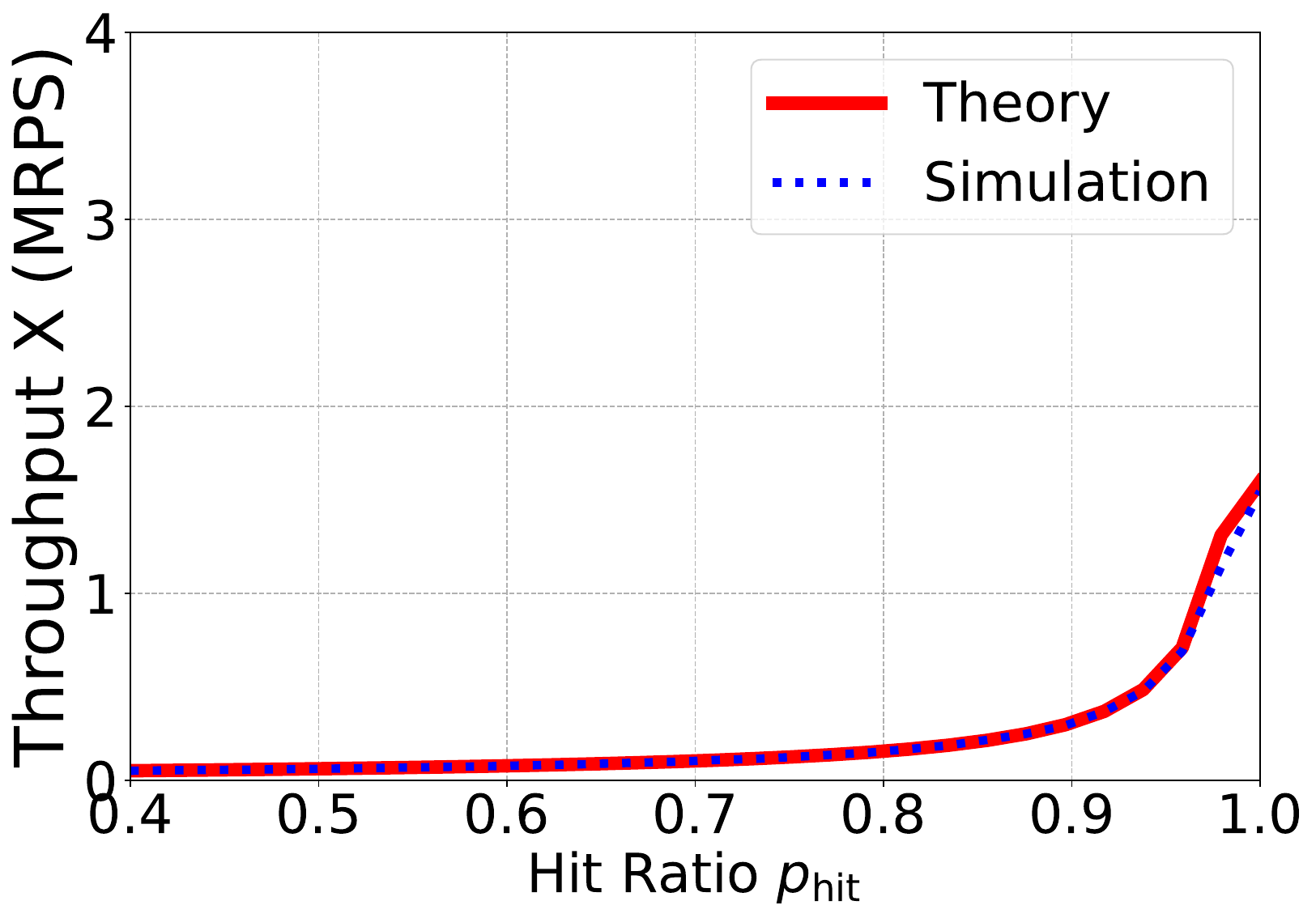}
  \caption{$L=500 \mu$s, MPL=16}

\end{subfigure}
\begin{subfigure}{.32\textwidth}
  \centering
  \includegraphics[height=1.1in]{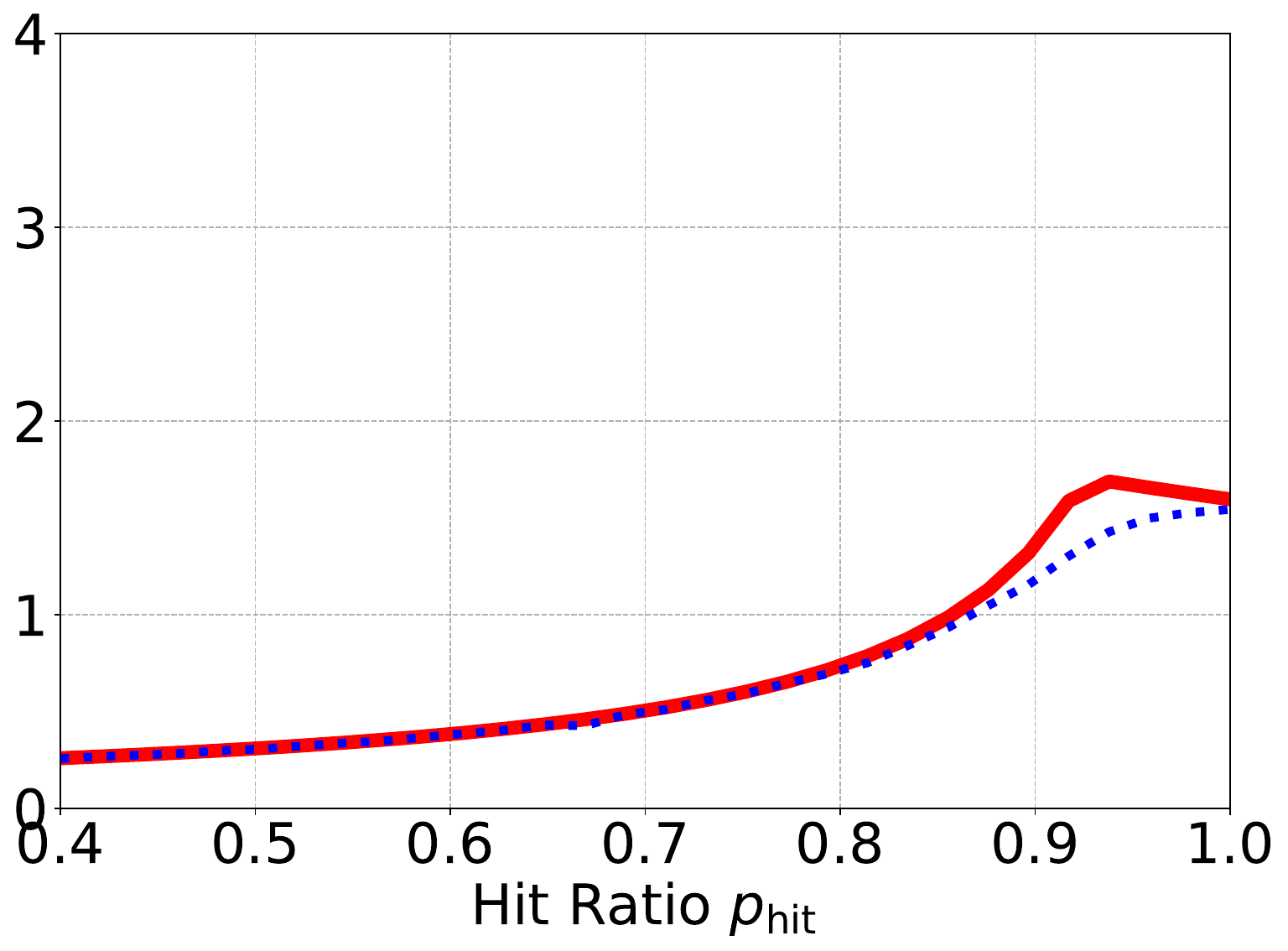}
  \caption{$L=100 \mu$s, MPL=16}

\end{subfigure}
\begin{subfigure}{.32\textwidth}
  \centering
  \includegraphics[height=1.1in]{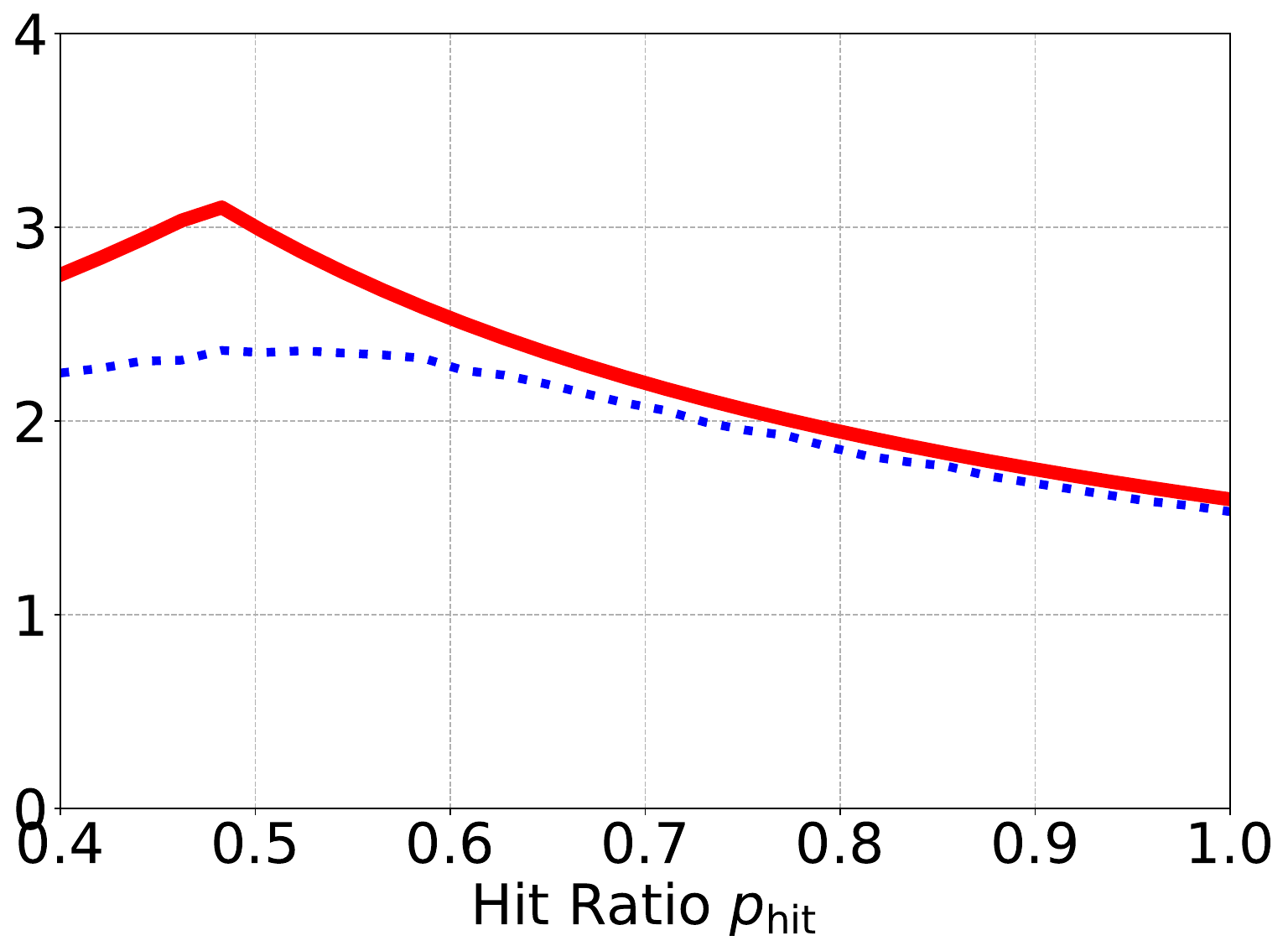}
  \caption{$L=5 \mu$s, MPL=16}

\end{subfigure}

\begin{subfigure}{.32\textwidth}
  \centering
  \includegraphics[height=1.1in]{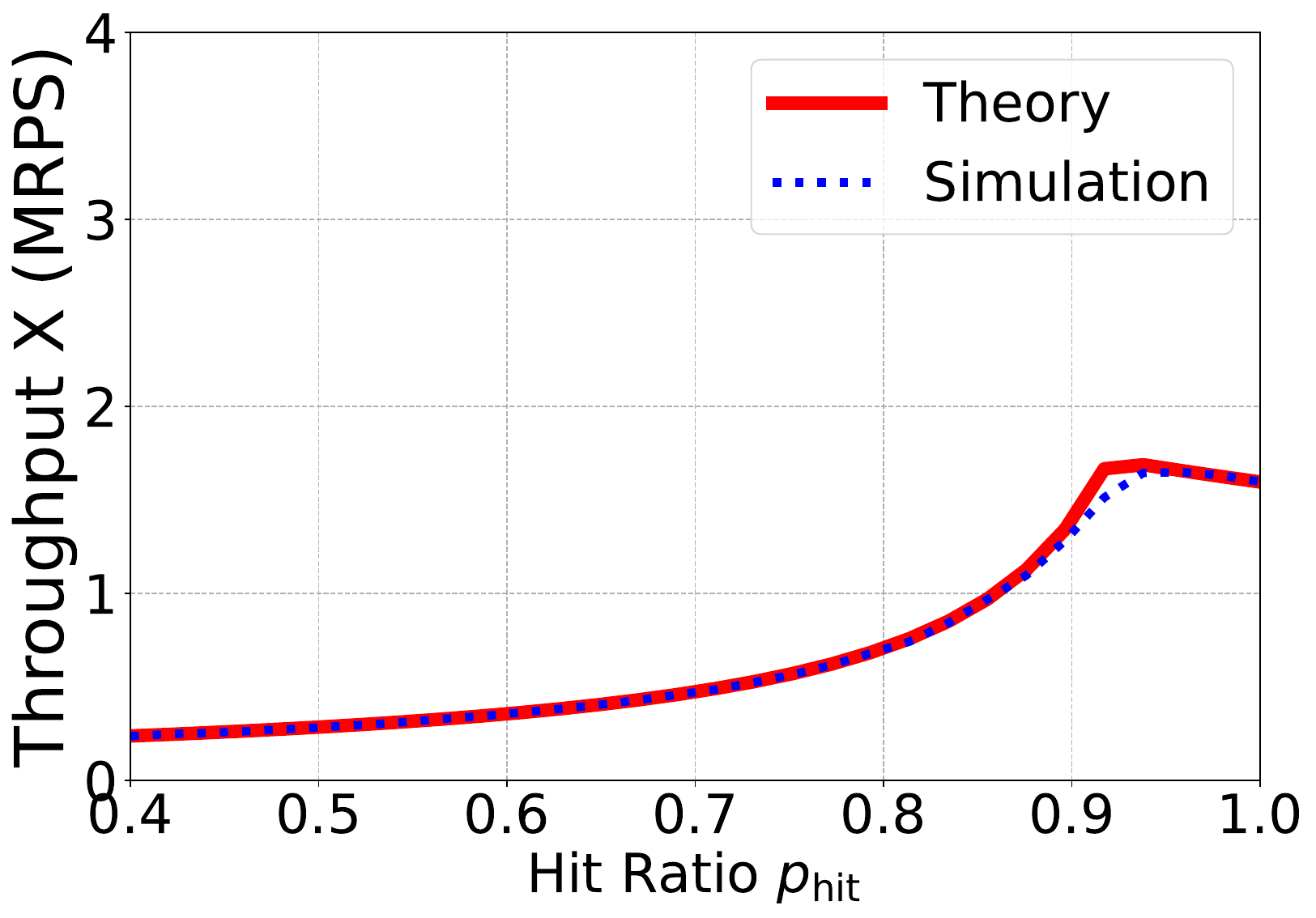}
  \caption{$L=500 \mu$s, MPL=72}
  \label{fig:intro:sub1}
\end{subfigure}
\begin{subfigure}{.32\textwidth}
  \centering
  \includegraphics[height=1.1in]{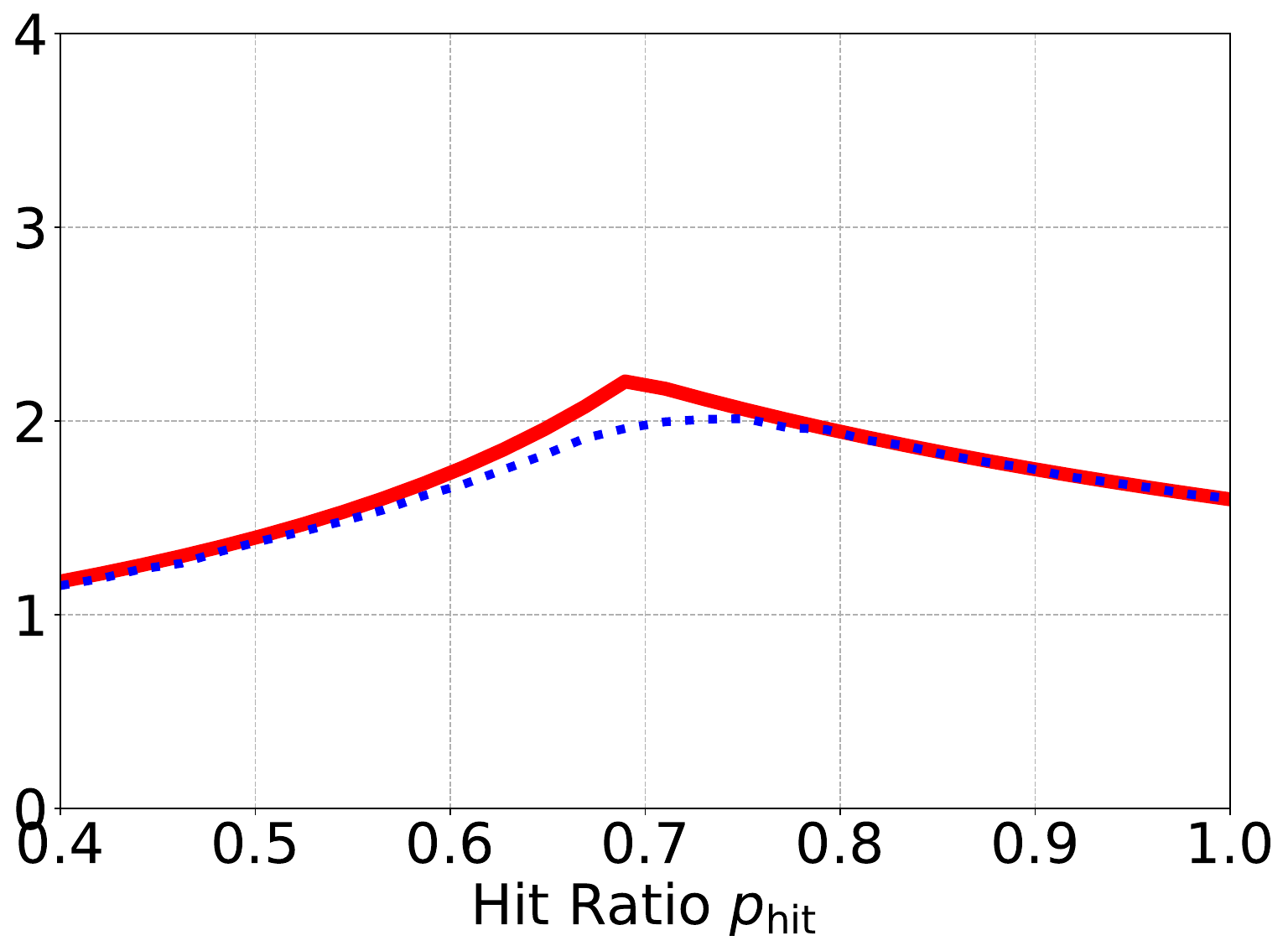}
  \caption{$L=100 \mu$s, MPL=72}
  \label{fig:intro:sub2}
\end{subfigure}
\begin{subfigure}{.32\textwidth}
  \centering
  \includegraphics[height=1.1in]{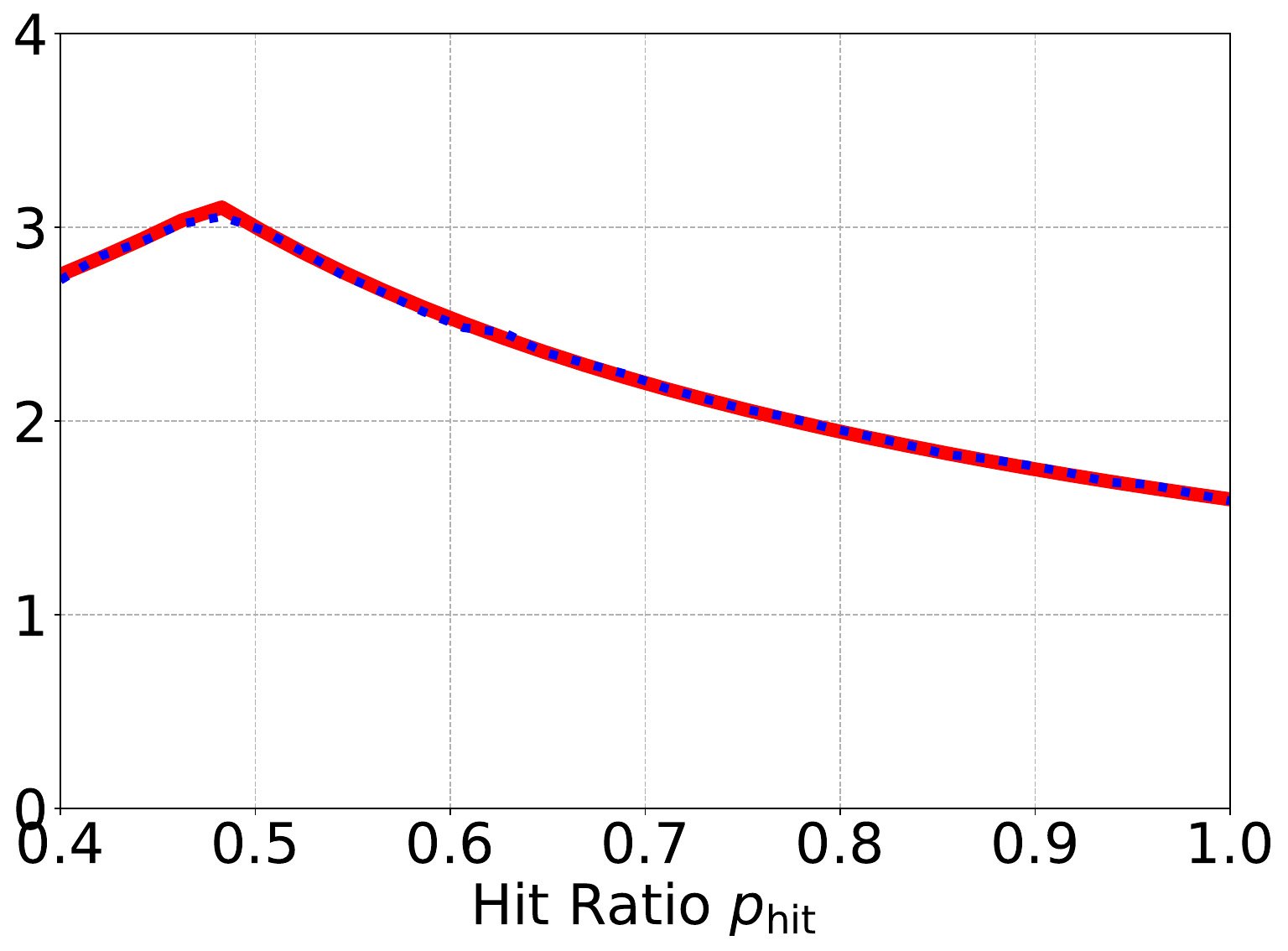}
  \caption{$L=5 \mu$s, MPL=72}
  \label{fig:intro:sub3}
\end{subfigure}

\begin{subfigure}{.32\textwidth}
  \centering
  \includegraphics[height=1.1in]{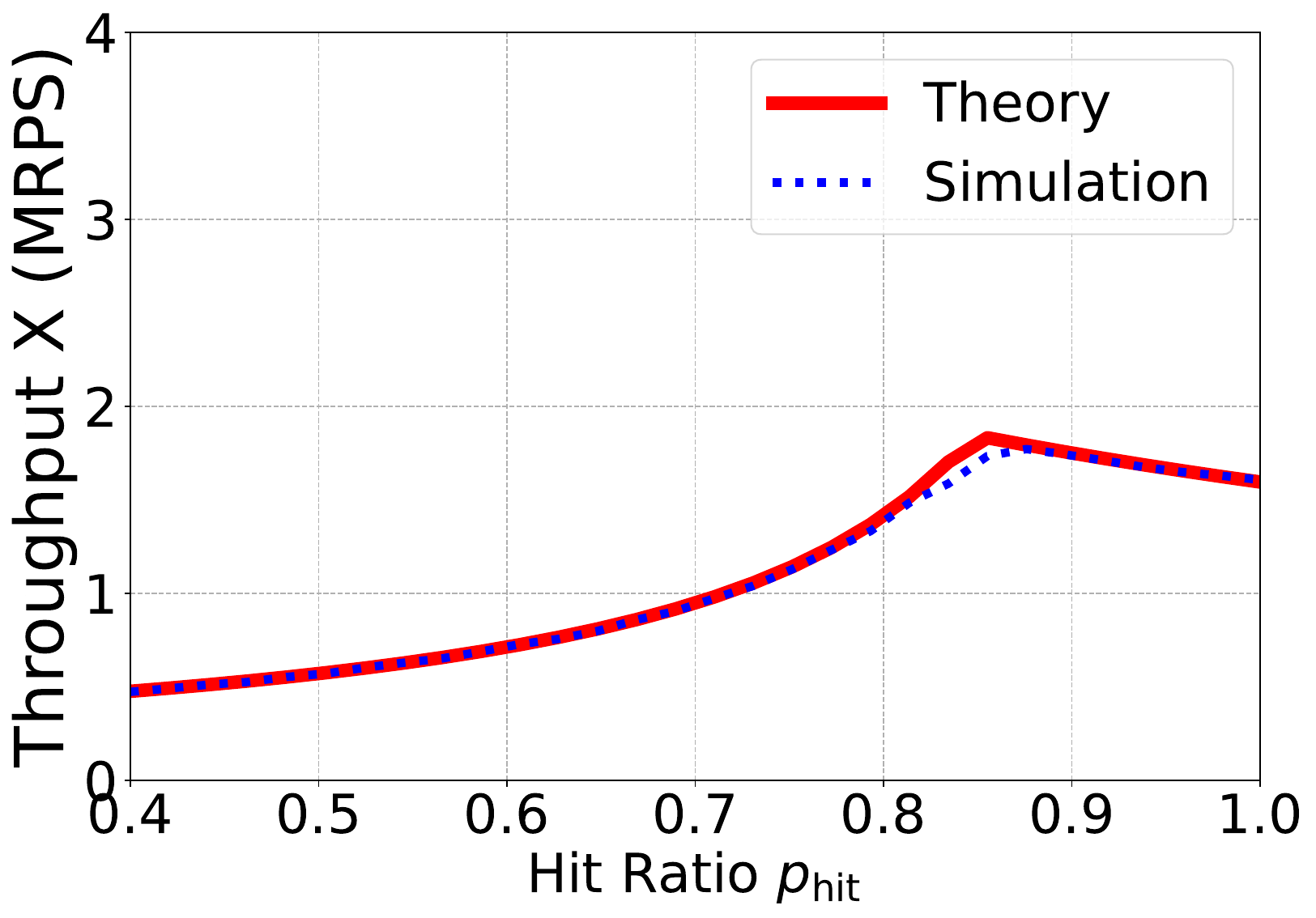}
  \caption{$L=500 \mu$s, MPL=144}
  \label{fig:intro:sub4}
\end{subfigure}%
\begin{subfigure}{.32\textwidth}
  \centering
  \includegraphics[height=1.1in]{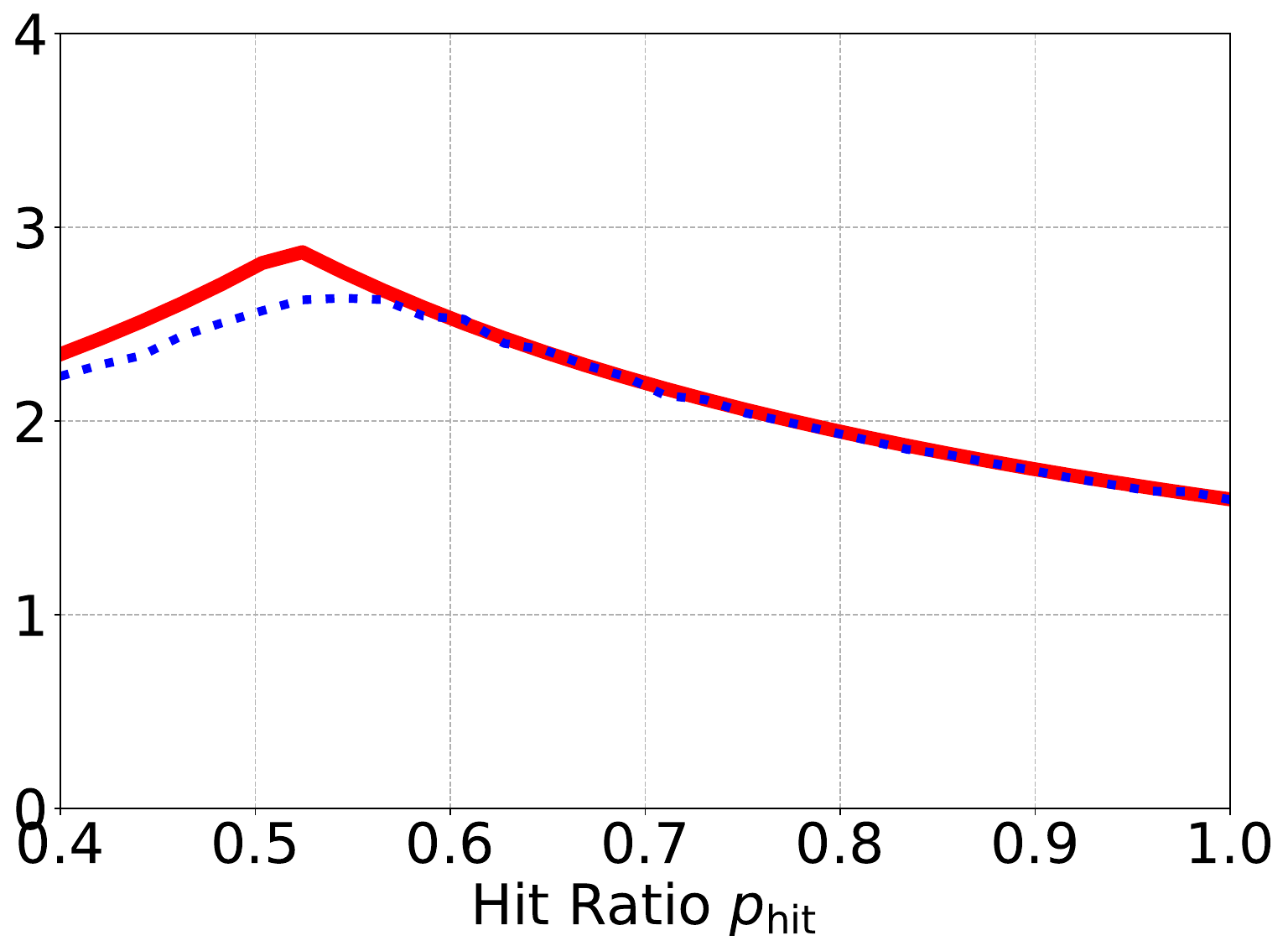}
  \caption{$L=100 \mu$s, MPL=144}
  \label{fig:intro:sub5}
\end{subfigure}
\begin{subfigure}{.32\textwidth}
  \centering
  \includegraphics[height=1.1in]{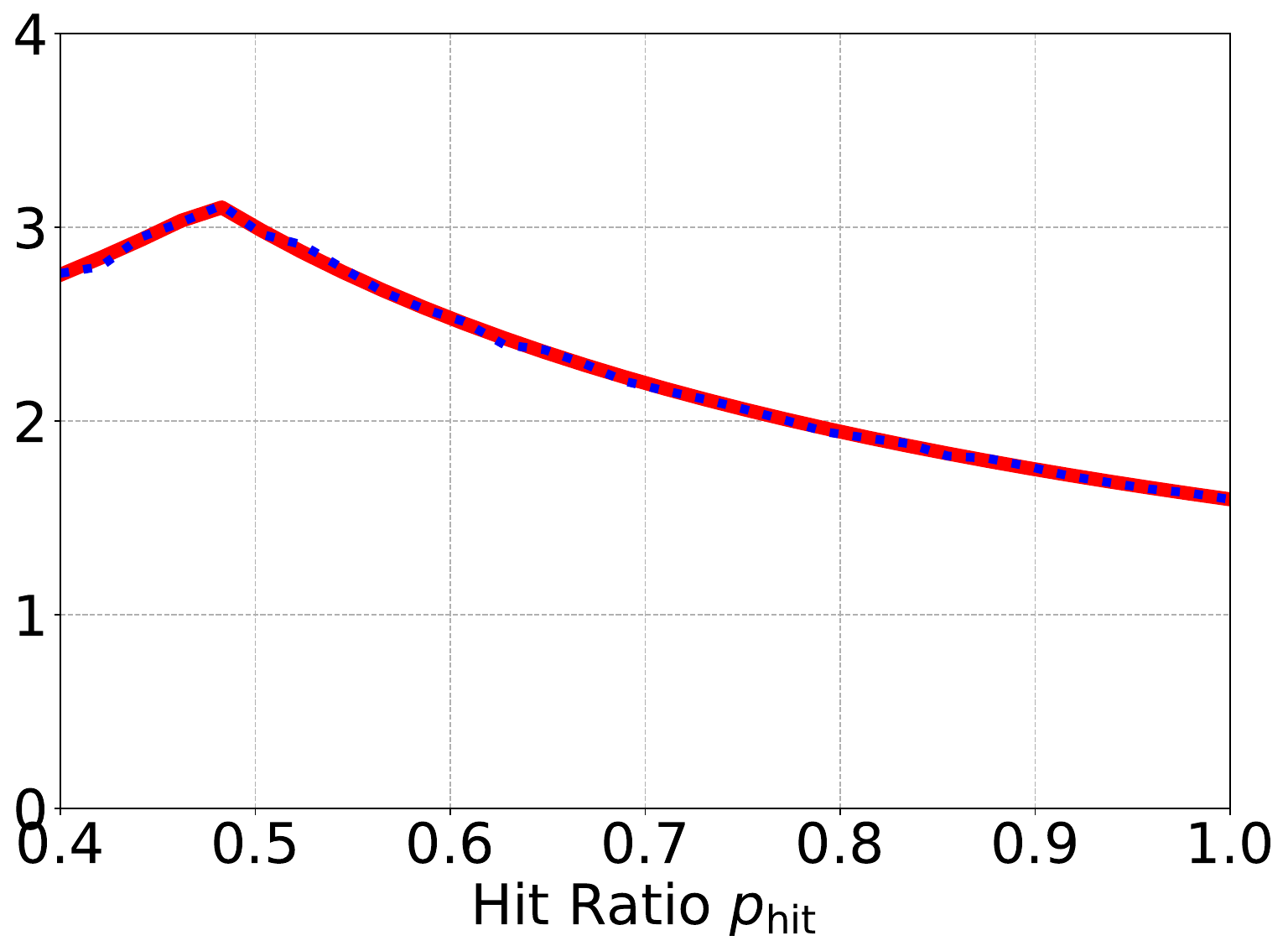}
  \caption{$L=5 \mu$s, MPL=144}
  \label{fig:intro:sub6}
\end{subfigure}
\vspace{-0.7em}
\caption{The throughput of a Segmented LRU cache is affected by the hit ratio, but also by the disk latency ($L$) and the MPL.\vspace{-1.6em}}
\label{fig:intro:slru-simu}
\end{figure*}

\noindent\textbf{Results of analysis and simulation.}
Figure~\ref{fig:intro:slru-simu} shows the results of analysis (red solid lines) and simulation (blue dotted lines) for SLRU. We have not implemented SLRU.

We also evaluate the effect of another trend, increasing the number of CPU cores, namely the Multi-Programming Level. The effect of {\em increasing the MPL} is shown in Figure~\ref{fig:intro:slru-simu} when looking from top (MPL = 16) to bottom (MPL = 144). We see that the point at which throughput starts to deteriorate, $p^*_{hit}$, moves earlier for higher MPL, as well as lower disk latency. This effect makes sense since a higher level of concurrency means that the delink operation becomes bottlenecked sooner.

Like LRU, SLRU also experiences reduced throughput with high hit ratios.
The queueing theory shows this mathematically.  More intuitively, the bottleneck here is often the delinkT operation.
Therefore, increasing hit ratio makes more requests queue behind the delinkT server which is already bottlenecked, hence lowering the throughput.
Note that this same behavior would happen if the DelinkB operation were the bottleneck.  Thus using a more complex policy, like SLRU, which has more queues, does not alleviate the bottleneck on the hit path.


\subsection{S3-FIFO}

S3-FIFO~\cite{S3FIFO} is a new algorithm that claims to have state-of-the-art hit ratio. 
We have not implemented this policy, but will study it via analysis and simulation.  S3-FIFO divides the global list of all elements in the cache into two global FIFO-ordered lists.  The first list is called the Small List, $S$, because it has a fixed short length, while the second list is called the Main List, $M$.  Typically, the $S$-List contains 10\% of the items while the $M$-List contains the rest (we used these numbers in our model).  S3-FIFO is very similar to CLOCK in that only a miss causes work to be done, while a hit only involves setting a bit.

There is also a Ghost element which determines whether the requested item should be stored in the $S$-List or the $M$-List.  The 
Ghost's decision is generally based on whether the object was accessed within the last $x$ misses of the cache, where $x$ is the number of items in the $M$-List.  Ghost lookup is similar in speed to a cache lookup, so $\E{Z_{ghost}} = 0.51 \mu s$.

As in CLOCK, every object in the cache has a special bit.
When the object first enters the cache, the object is assigned a bit of $0$.  If the object is accessed while it is in the cache, then its bit is changed from $0$ to $1$.  

When an object $d$ is requested, if $d$ is in the cache (regardless of which list), we get $d$ and set $d$'s bit to $1$ (if it's not already $1$), completing the request.  
If, on the other hand, $d$ is not in the cache (a ``miss"), then we need to get $d$ from disk.  We now use the Ghost to figure out in which list to store $d$.  If $d$ was ``missed recently" (meaning it was missed sometime within the last $x$ misses of the cache), then we store $d$ in the $M$-List.  Otherwise, we store $d$ in the $S$-List.  Since both are FIFO lists, it is advantageous to $d$ to be stored in the $M$-List, which is a lot longer.  Either way, we still need a Head Update (to append $d$) and also a Tail Update (to remove the object at the tail), thus keeping both lists at their original size.  Before the object at the tail of the $S$-List is thrown out, it has the opportunity to move to the $M$-List.  This happens if and only if its bit is $1$.  Otherwise it is tossed.
Objects in the $M$-List never move to the $S$-List.  

\begin{figure}[t]
\centerline{\psfig{file=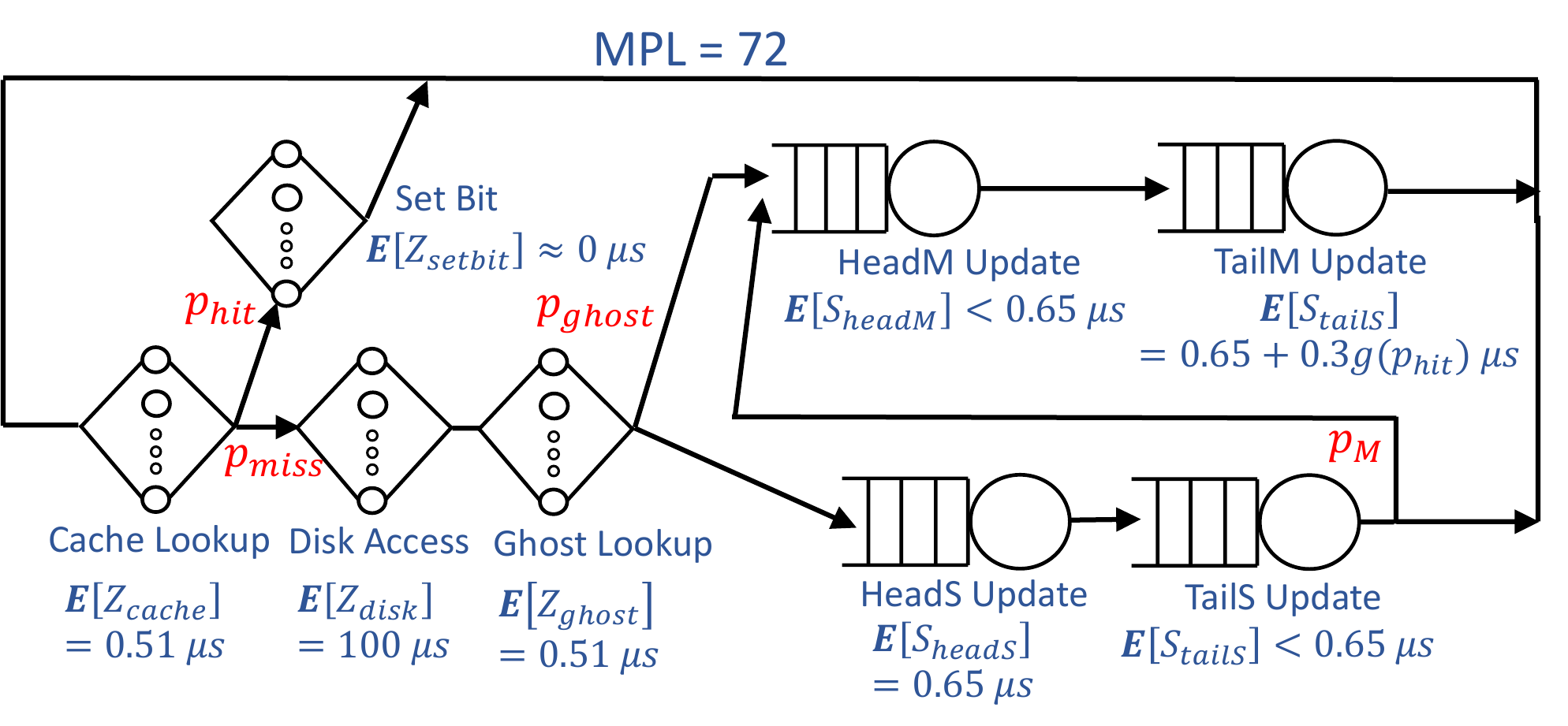,height=1.6in}}
\vspace{-1.0em}
\caption{Queueing model of S3-FIFO cache. \vspace{-1.1em}}
\label{fig:S3FIFOcache}
\end{figure}

To do a queueing analysis of the network in Figure~\ref{fig:S3FIFOcache}, we need three more things:
First, we need to know the service times for all  potential bottleneck operations.  These are the same as the numbers in the CLOCK network.   Second, and more challenging, we need to understand the fraction of requests that the Ghost sends to the $M$ list (denoted $p_{ghost}$) as opposed to the $S$ list. 
Third, we need to know the fraction of items, $p_M$, at the tail of the $S$ list that have a $1$ bit associated with them as opposed to a $0$ bit.  
We used our simulation to estimate both $p_{ghost}$ and $p_M$ as functions of $p_{hit}$.
{\color{chighlight}
Our results are expressible in terms of the $\chi^2$ function, which is denoted by $h(x; a, b,c)$, where $\Gamma(\cdot)$ is the Gamma factorial function. We were able to match our measurements of $p_{ghost}$ and $p_{M}$ to the expressions below, which involve both $p_{hit}$ and the  $\chi^2$ function with particular parameters: 

\begin{eqnarray*}
    h(x; a, b, c) &= & \frac{1}{2^{\frac{a}{2}} \Gamma\left(\frac{a}{2}\right) c^{a}} \left(\frac{x - b}{c}\right)^{\frac{a}{2} - 1} e^{-\frac{x - b}{2 c}}\\
    p_{ghost} &=& \frac{h(65 \cdot (1-p_{hit}); 4.4912, 1.1394, 3.595)}{1-p_{hit}}\\
    p_{M} &=& \frac{h(400 \cdot (1-p_{hit}); 2.2870, 4.5309, 26.5874)}{1-p_{hit}}\\
\end{eqnarray*}

\subsubsection*{Analysis of the queueing network model}  

The {\em mean think time} of the system is:
\begin{eqnarray*}
    \E{Z} & = & \E{Z_{cache}} + p_{miss} \cdot (\E{Z_{disk}} + \E{Z_{ghost}}) \\
    & = & 0.51 + p_{miss} \cdot 100.51 = 101.02 - 100.51 p_{hit}
\end{eqnarray*}
For each queue, we now compute the {\em device demand}: 

Let $q_{ghost} = 1 - p_{ghost}$.  Then, 
\begin{eqnarray*}
D_{tailS} &<& D_{headS} = p_{miss}  \cdot (1 - p_{ghost} ) \cdot 0.65 \\
D_{headM} &<& (p_{miss} \cdot q_{ghost} \cdot p_{M} + p_{miss} \cdot p_{ghost}) \cdot 0.65 \\
D_{tailM} &=& (p_{miss} \cdot q_{ghost}\cdot p_{M} + p_{miss} \cdot p_{ghost}) \cdot g(p_{hit})\\
\end{eqnarray*}
    
\vspace{-0.4em}
Because $D_{tailM} \geq D_{headM}$, the bottleneck device can be the tail update on the Main List or the head update on the Small List, based on $p_M$, so:
  $$D_{max} = \max\left(D_{tailM}, D_{headS}\right). $$

The {\em total demand}, $D$, is the sum of the device demands: 
\begin{eqnarray*}
  D & = &  D_{headS} + D_{tailS} + D_{headM} + D_{tailM}  \\
\end{eqnarray*}
Because $D$ now has four different components, it becomes messy to write out the explicit upper and lower bounds on $D$.  Nonetheless, we have computed upper bounds on throughput, $X$, via the usual formula from \cite[Theorem 7.1]{PerformanceModeling13}.
}

\begin{figure*}[t]
\centering
\begin{subfigure}{.33\textwidth}
  \centering
  \includegraphics[height=1.1in]{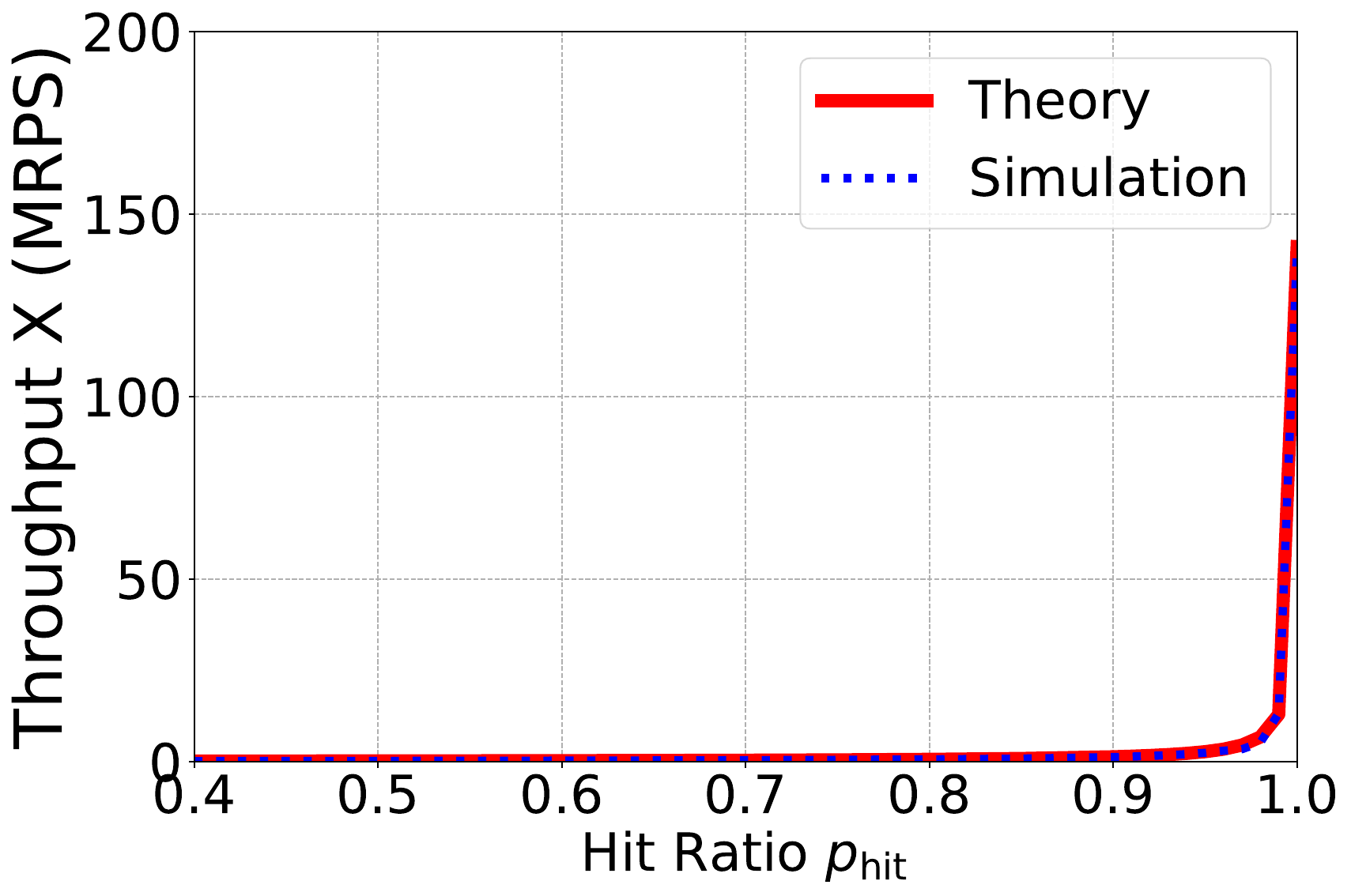}
  \caption{Disk latency 500 $\mu$s}
\end{subfigure}%
\begin{subfigure}{.33\textwidth}
  \centering
  \includegraphics[height=1.1in]{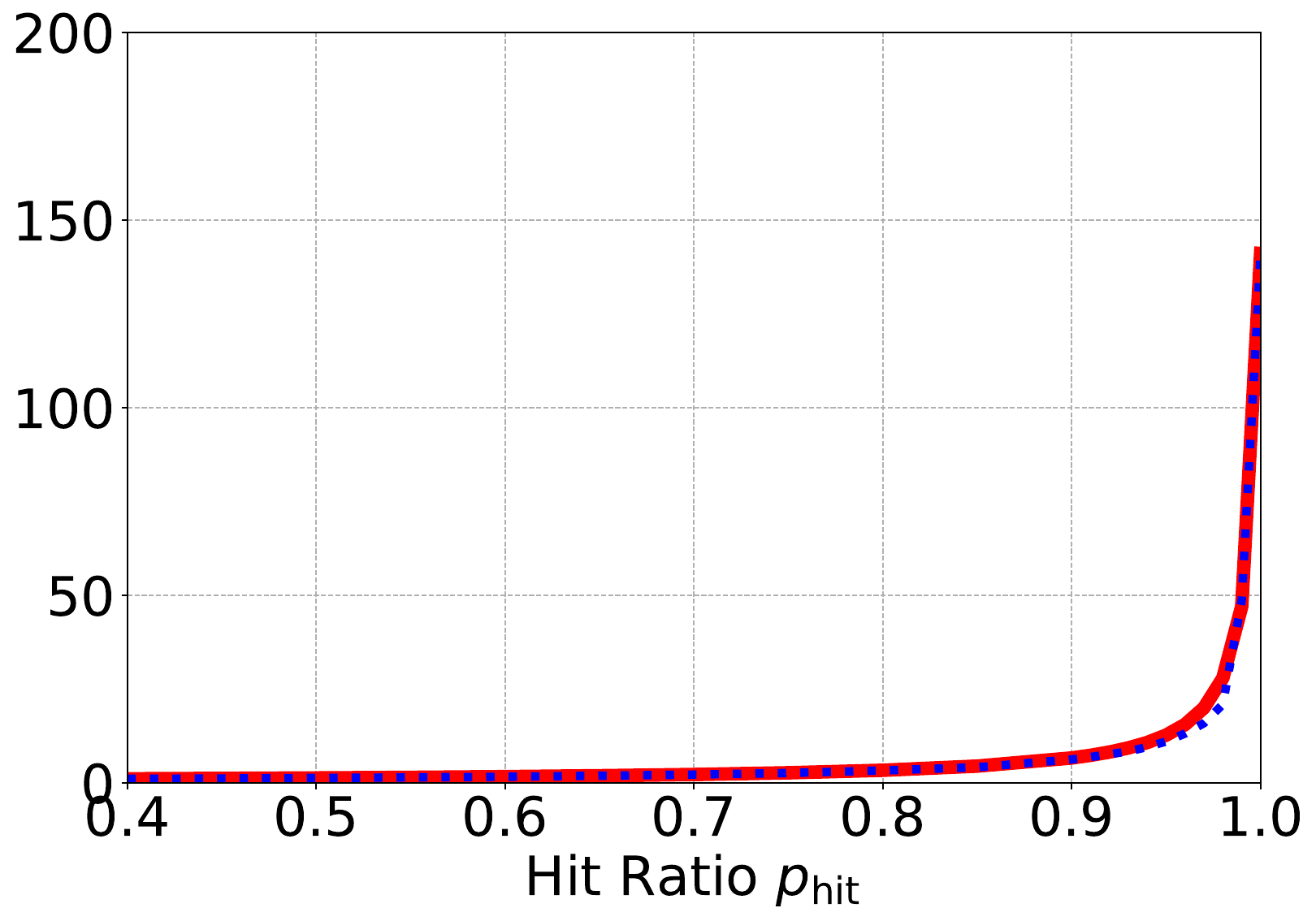}
  \caption{Disk latency 100 $\mu$s}
\end{subfigure}
\begin{subfigure}{.33\textwidth}
  \centering
  \includegraphics[height=1.1in]{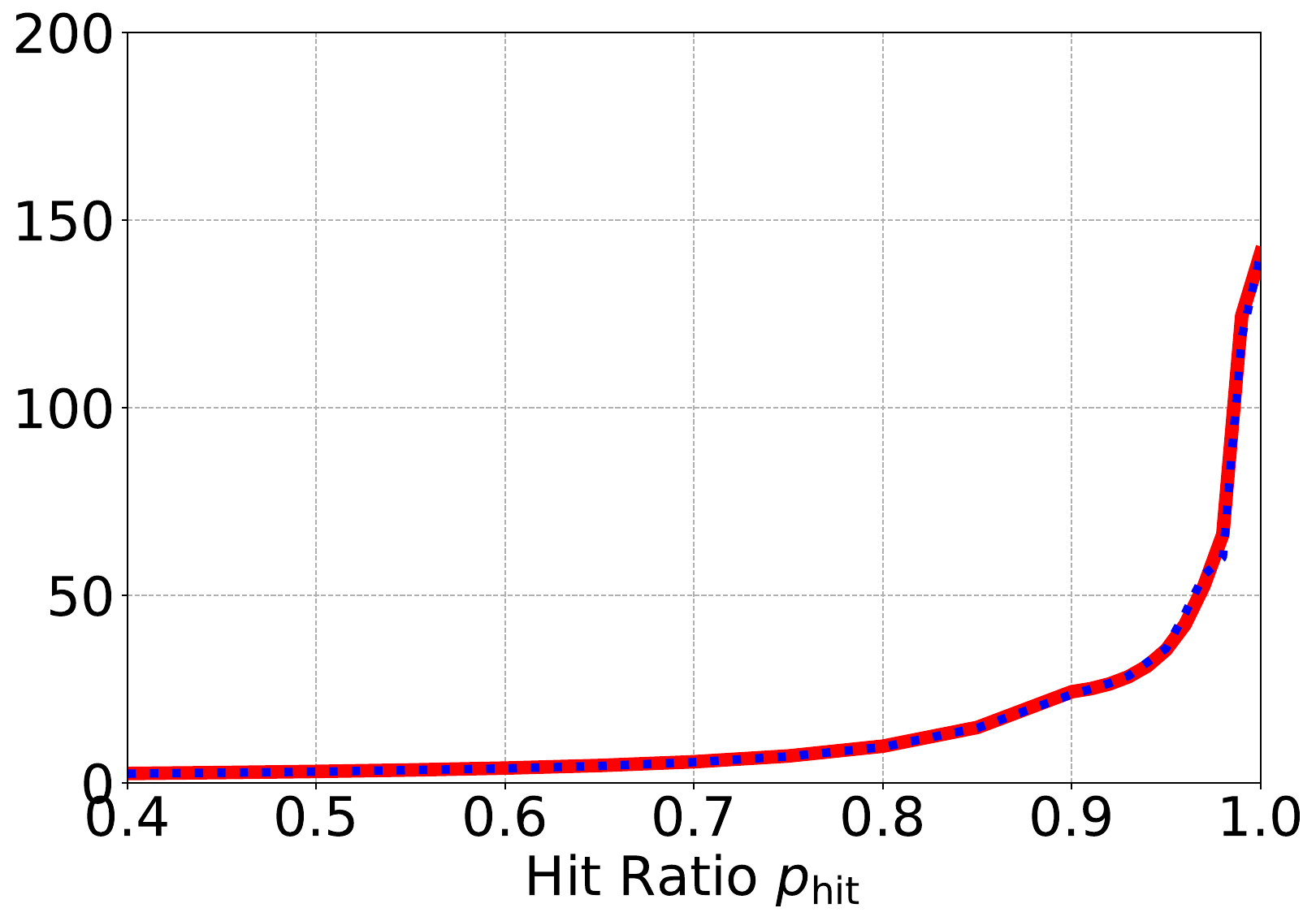}
  \caption{Disk latency 5 $\mu$s}
\end{subfigure}
\vspace{-0.8em}
\caption{Results for theory and simulation under an S3-FIFO cache.   Throughput of the S3-FIFO cache always increases with the hit ratio, for all disk latencies. \vspace{-1.6em}}
\label{fig:s3fifo-simu}
\end{figure*}

\vspace{.1in}
\noindent\textbf{Results of analysis and simulation. }
Figure~\ref{fig:s3fifo-simu} shows the results of analysis (red solid) and simulation (blue dotted) for S3-FIFO.
We see that increasing the hit ratio, $p_{hit}$, always leads to higher throughput, regardless of the mean disk latency.   This holds because 
the bottleneck device (no matter whether it is the headS update or tailM update) is in the {\em miss } path, not in the hit path. Hence, increasing $p_{hit}$ does not increase demand on the bottleneck device.  

\section{Discussion}
\label{s:discussion}





\subsection{A classification of eviction algorithms}
\label{s:discussion:summary}
Throughout our paper, we find that common eviction algorithms can be classified into two categories: {\bf LRU-like} algorithms and {\bf FIFO-like} algorithms. The throughput of LRU-like algorithms (LRU, Probabilistic LRU with lower $q$, and SLRU) typically drops when the hit ratio becomes high.  By contrast, the throughput of FIFO-like algorithms (FIFO, Probabilistic LRU with very high $q$, CLOCK, and S3-FIFO) always increases with the hit ratio. 


\vspace{-2.0em}
\begin{table}[h]
\centering
\caption{Conjectured classification of additional algorithms.}
\label{tab:alg2}
\begin{tabular}{@{}ll@{}}
\toprule
LRU like           & {\small ARC~\cite{megiddo_arc_2003}, LIRS~\cite{jiang_lirs_2002}, TinyLFU~\cite{einziger_tinylfu_2017},} \\ & {\small LeCaR~\cite{vietri_driving_2018}, CACHEUS~\cite{rodriguez_learning_2021}, LFU} \\
\hline
FIFO like          & {\small CLOCK variants~\cite{carr_wsclock_1981}, SIEVE~\cite{sieve}, QDLP~\cite{yang_fifo_2023}, }\\ & {\small Hyperbolic~\cite{blankstein_hyperbolic_2017}, Random, LHD~\cite{beckmann_lhd_2018}, LRB~\cite{song_learning_2020} }  \\ 
\bottomrule
\vspace{-2.0em}
\end{tabular}
\end{table}

In Table~\ref{tab:alg2}, we conjecture the behavior of other algorithms based on what we've learned. 
The LRU-like algorithms all have the common feature that they perform a delink operation upon a cache hit.  This delink operation becomes the bottleneck, meaning that increasing the hit ratio will increase the queueing time at the delink queue, which will lead to lower throughput.  
The FIFO-like algorithms  have the common feature that they do not update the global data structure upon cache hits.  Consequently, increasing $p_{hit}$ will not increase the queue length at the bottleneck device, and therefore will not lead to lower throughput.  

These FIFO-like algorithms can be further subdivided into two types. Algorithms in the first type use FIFO as the basic building block, e.g., CLOCK variants~\cite{carr_wsclock_1981}, S3-FIFO~\cite{S3FIFO}, QDLP~\cite{yang_fifo_2023}, and SIEVE~\cite{sieve}. 
Algorithms in the second type do not have a global data structure, but rather uses random sampling to choose eviction candidates. Examples include LHD~\cite{beckmann_lhd_2018}, LRB~\cite{song_learning_2020}, and Hyperbolic~\cite{blankstein_hyperbolic_2017}. 
As these algorithms do not maintain a global data structure, they are never bottlenecked by cache hits, so throughput only increases with $p_{hit}$.

\subsection{Improving future caching systems}

Future CPUs will have more cores, further increasing their concurrency.
At the same time, disks are getting faster, with 
high-end disks having single-digit microseconds of latency. 
Figure~\ref{fig:intro:slru-simu} shows the effects of these two trends.
 We see that the critical hit ratio, $p^*_{hit}$, after which throughput drops, will shift to smaller values in the future.  
Therefore, fixing the problems of LRU-like algorithms, which are the predominant caching algorithms used today, will become more important.  

The easiest mitigation is to reduce the number of delink operations. This can be achieved by performing the delink operation probabilistically (probabilistic LRU).
As we have shown in Section~\ref{s:models}, when $q$ is very high, probabilistic LRU behaves like FIFO, where increasing the hit ratio only helps. This is the high-level idea behind a very recent paper, \cite{qiu_frozenhot_2023}.

Another approach is to send some of the requests to the disk directly, bypassing the cache, when cache load is high. 
We simulated this solution and found that throughput stays constant after the critical $p^*_{hit}$ point, rather than dropping.   

One might be thinking at this point that, in a future with higher concurrency and faster disks, one might want to forgo  LRU altogether, in favor of FIFO.   However there are reasons why cache designers prefer LRU to FIFO.  Specifically FIFO is less efficient in its use of cache space --  a larger cache is needed to obtain the same hit ratio under FIFO as under LRU.  Thus, what we believe will be used in the future is some {\em combination} of FIFO and LRU which looks a lot like FIFO, but still maintains the efficiency of LRU.  There are a couple very recent papers (2023) which go in this direction~\cite{S3FIFO, sieve}. 
 




\section{Prior Work}

DRAM-based caches are increasingly deployed in today's system stack and are widely studied in research~\cite{berger_adaptsize_2017,kirilin_rl-cache_2019,yang_cachesack_2022,berger_towards_2018,eisenman_flashield_2019,gill_amp_2007,li_c-miner_2004,plonka_context-aware_2008,oneil_lru-k_1993,vietri_driving_2018,rodriguez_learning_2021,song_halp_2023,yang_gl-cache_2023,blankstein_hyperbolic_2017,azim_recache_2017,akhtar_avic_2019,cheng_take_2023}.

Most of the above works use the {\em hit ratio} as a proxy for system performance.  Those that do look at throughput only consider it a single thread (i.e., MPL=1), e.g., DistCache~\cite{liu_distcache_2019}, p-redis~\cite{pan_predis_2019}, LRB~\cite{song_learning_2020}, GL-Cache~\cite{yang_gl-cache_2023}. By contrast, our paper models modern systems with MPL = 72 concurrent threads. 

Two papers deserve a bit more attention because, like our paper, both are aimed at increasing throughput and both assume a high number of threads.  In the first paper, FrozenHot~\cite{qiu_frozenhot_2023}, the authors fix the hit ratio at $99\%$ and do not study throughput as a function of changing hit ratio.  
The second paper, NHC~\cite{wu_storage_2021}, is quite different from ours. This paper uses NVM and Optane caches, which have much lower concurrency.   Their bottlenecks are thus far different from ours; in particular the cache operations are not modeled whatsoever.
Finally, neither the FrozenHot nor NHC papers have any analytic modeling component.

\section{Conclusion}
\label{s:conclusion}


This paper examines the effect of the cache hit ratio on the request throughput of DRAM-based software caching systems.  
We use a {\em three-pronged approach}. First we create a queueing model of the particular eviction policy, which we evaluate analytically to obtain an upper bound on the throughput as a function of the hit ratio. We next use simulation to exactly evaluate the queueing network. Finally we  study the eviction policy in implementation.

For all policies studied, our simulation and implementation agree within 5\%, indicating that our queueing network models are quite accurate. Another finding is that our analysis, while only providing an upper bound, is an excellent indicator of $p^*_{hit}$, the critical hit ratio at which throughput starts decreasing. The takeaway is that, for the evaluation of future eviction policies, it suffices to use analytic upper bounds to understand whether increasing hit ratio will help.



Our queueing analysis illucidates why throughput drops with higher hit ratio for LRU-like algorithms.  In a nutshell, LRU-like algorithms require updating the global linked list upon a cache hit. When the hit ratio becomes very high, this update operation becomes the system bottleneck, meaning that many requests queue up there.  As a consequence, increasing the hit ratio increases the queueing time at this bottleneck, which reduces system throughput.   
Based on this intuition, we conjecture that many algorithms that we have not yet studied (ARC, LIRS, LFU, CACHEUS, LeCAR) should also exhibit this perverse behavior.  

We also study two major trends in computing and storage: newer CPUs have an increasing number of cores, while disk latencies are steadily decreasing.  We show that both these two trends imply that LRU-like algorithms will only behave worse in the future; i.e., throughput will start to decrease at lower and lower values of the 
hit ratio.  We offer some suggestions to remedy this problem.




\bibliographystyle{plain}
\bibliography{references_jason, other}


\end{document}